\documentclass[traditabstract]{aa} 

\usepackage{graphicx, times, txfonts}

\newcommand{\less}{\raisebox{-1.1mm}{$\stackrel{<}{\sim}$}} 
\newcommand{\more}{\raisebox{-1.1mm}{$\stackrel{>}{\sim}$}} 
\newcommand{\msol}{\mbox{M$_{\odot}$}} 
 
\newcommand{\msolyr}{{M$_{\odot}$}\,yr$^{-1}$}

\newcommand{\ks}{km s$^{-1}$}

\begin{document}

\title{
MESS (Mass-loss of Evolved StarS), a Herschel Key 
Program\thanks{Herschel is an ESA space observatory with science
  instruments provided by European-led Principal Investigator
  consortia and with important participation from NASA.}  }
 
\author{ 
M.A.T.~Groenewegen\inst{1}\and 
C.~Waelkens\inst{2}\and
M.J.~Barlow\inst{3}\and
F.~Kerschbaum\inst{4}\and
P.~Garcia-Lario\inst{5}\and
J.~Cernicharo\inst{6}\and
J.A.D.L.~Blommaert\inst{2}\and
J.~Bouwman\inst{7}\and
M.~Cohen\inst{8}\and
N.~Cox\inst{2}\and
L.~Decin\inst{2,9}\and
K.~Exter\inst{2}\and
W.K.~Gear\inst{10}\and
H.L.~Gomez\inst{10}\and
P.C.~Hargrave\inst{10}\and
Th.~Henning\inst{7}\and 
D.~Hutsem\'ekers\inst{15}\and
R.J.~Ivison\inst{11}\and
A.~Jorissen\inst{16}\and
O.~Krause\inst{7}\and
D.~Ladjal\inst{2}\and
S.J.~Leeks\inst{12}\and
T.L.~Lim\inst{12}\and
M.~Matsuura\inst{3,18}\and
Y.~Naz\'e\inst{15}\and
G.~Olofsson\inst{13}\and
R.~Ottensamer\inst{4,19}\and
E.~Polehampton\inst{12,17}\and
T.~Posch\inst{4}\and
G.~Rauw\inst{15}\and
P.~Royer\inst{2}\and
B.~Sibthorpe\inst{7}\and
B.M.~Swinyard\inst{12}\and
T.~Ueta\inst{14}\and
C.~Vamvatira-Nakou\inst{15}\and
B.~Vandenbussche\inst{2}\and
G.C.~Van~de~Steene\inst{1}\and
S.~Van~Eck\inst{16}\and
P.A.M.~van~Hoof\inst{1}\and
H.~Van~Winckel\inst{2}\and
E.~Verdugo\inst{5}\and
R.~Wesson\inst{3}
}

\institute{ 
Koninklijke Sterrenwacht van Belgi\"e, Ringlaan 3, B--1180 Brussel, Belgium 
\and
Institute of Astronomy, University of Leuven, Celestijnenlaan 200D, B--3001 Leuven, Belgium  
\and
Department of Physics and Astronomy, University College London, Gower Street, London WC1E 6BT
\and
University of Vienna, Department of Astronomy, T\"urkenschanzstrasse 17, A--1180 Wien, Austria 
\and
Herschel Science Centre, European Space Astronomy Centre, Villafranca del Castillo. 
Apartado de Correos 78, E--28080 Madrid, Spain 
\and
Astrophysics Dept, CAB (INTA-CSIC), Crta Ajalvir km4, 28805 Torrejon de Ardoz, Madrid, Spain 
\and
Max-Planck-Institut f\"ur Astronomie, K\"onigstuhl 17, D--69117 Heidelberg, Germany 
\and
Radio Astronomy Laboratory, University of California at Berkeley, CA 94720, USA 
\and
Sterrenkundig Instituut Anton Pannekoek, University of Amsterdam, 
Kruislaan 403, NL--1098 Amsterdam, The Netherlands
\and
School of Physics and Astronomy, Cardiff University, 5 The Parade, Cardiff, Wales CF24 3YB, UK  
\and
UK Astronomy Technology Centre, Royal Observatory Edinburgh, Blackford Hill, Edinburgh EH9 3HJ, UK 
\and
Space Science and Technology Department, Rutherford Appleton Laboratory, Oxfordshire, OX11 0QX, UK  
\and
Dept of Astronomy, Stockholm University, AlbaNova University Center, 
Roslagstullsbacken 21, 10691 Stockholm, Sweden             
\and
Dept. of Physics and Astronomy, University of Denver, Mail Stop 6900, Denver, CO 80208, USA  
\and
Institut d'Astrophysique et de G\'eophysique, All\'ee du 6 ao\^{u}t, 17 - B\^{a}t. B5c B--4000 Li\`ege 1, Belgium
\and
Institut d'Astronomie et d'Astrophysique, Universit\'e libre de Bruxelles, CP 226, 
Boulevard du Triomphe, B-1050 Bruxelles, Belgium 
\and
Institute for Space Imaging Science,  University of Lethbridge, 
Lethbridge, Alberta,  T1J 1B1, Canada               
\and
Mullard Space Science Laboratory, University College London, Holmbury St. Mary, Dorking, Surrey RH5 6NT, United Kingdom  
\and
TU Graz, Institute for Computer Graphics and Vision, Inffeldgasse 16/II, A--8010 Graz, Austria 
} 
 
\date{received: 2010,  accepted: 2010} 
 
 
 
\abstract{
MESS (Mass-loss of Evolved StarS) is a Guaranteed Time Key Program
that uses the PACS and SPIRE instruments on board the \it Herschel Space Observatory \rm 
to observe a representative sample of evolved stars, that include asymptotic giant branch (AGB)
and post-AGB stars, planetary nebulae and red supergiants, as well as 
luminous blue variables, Wolf-Rayet stars and supernova remnants. 
In total, of order 150 objects are observed in imaging and about 50 objects in spectroscopy.

This paper describes the target selection and target list, and the observing strategy. 
Key science projects are described, and illustrated using results obtained during
\it Herschel\rm's science demonstration phase.
Aperture photometry is given for the 70 AGB and post-AGB stars 
observed up to October 17, 2010, which constitutes the largest single uniform database of 
far-IR and sub-mm fluxes for late-type stars.

}


\maketitle

\section{Introduction} 

Mass-loss is {\it the} dominating factor in the post-main sequence
evolution of almost all stars. For low- and intermediate mass stars
(initial mass \less 8 \msol) this takes place mainly on the
thermally-pulsing AGB (asymptotic giant branch) in a slow (typically
5-25 \ks) dust driven wind with large mass loss rates (up to 10$^{-4}$
\msolyr, see the contributions in the book edited by Habing \&
Olofsson 2003), which is also the driving mechanism for the slightly
more massive stars in the Red Supergiant (RSG) phase, while for
massive stars (initial mass \more 15 \msol) the mass loss takes place
in a fast (hundreds to a few thousand \ks) wind driven by radiation
pressure on lines at a moderate rate of a few 10$^{-6}$ \msolyr\ (Puls
et al. 2008).

Although mass loss is such an important process and has been studied
since the late 1960's with the advent of infrared astronomy, many
basic questions remain unanswered even after important missions such
as {\it IRAS} (Neugebauer et al. 1984), {\it ISO} (Kessler et al. 1996), 
{\it Spitzer} (Werner et al. 2004) and {\it AKARI}
(Murakami et al. 2007): what is the time evolution of the mass-loss rate, 
what is the geometry of the mass-loss process and how does this
influence the shaping of the nebulae seen around the central stars of
Planetary Nebulae (PNe) and Luminous Blue Variables (LBVs), 
can we understand the interaction of these winds with the interstellar 
medium (ISM) as initially seen by {\it IRAS} (e.g. Stencel et al. 1988) and 
confirmed by {\it AKARI} (Ueta et al. 2006, 2008) and {\it Spitzer} (Wareing et al. 2006),
what kind
of dust species are formed at exactly what location in the wind, 
what are the physical and chemical processes involved in driving the mass-loss
itself and how do they depend on the chemical composition of the photospheres?
With its improved spatial resolution compared to {\it ISO} and {\it  Spitzer}, 
larger field-of-view, better sensitivity, the extension to longer and
unexplored wavelength regions, and medium resolution spectrometers,
the combination of the Photodetector Array Camera and Spectrometer (PACS, Poglitsch et al. 2010) 
and the Spectral and Photometric Imaging Receiver (SPIRE,  Griffin et al. 2010) 
observations on board the {\it Herschel Space Observatory}
(Pilbratt et al. 2010) have the potential to lead to a significant
improvement in our understanding of the mass-loss phenomenon.
This is not only important for a more complete understanding of these
evolutionary phases {\it per se}, but has potentially important
implications for our understanding of the life cycle of dust and gas in the universe.

Dust is not only present and directly observable in our Galaxy and
nearby systems like the Magellanic Clouds, but is already abundantly
present at very early times in the universe, e.g. in damped
Lyman-alpha systems (Pettini et al. 1994), sub-millimetre selected galaxies 
(Smail et al. 1997) and high-redshift quasars (e.g. Omont et al. 2001, Isaak et al. 2002).
The inferred far-IR (FIR) luminosities of samples of $5 < z < 6.4$
quasars are consistent with thermal emission from warm dust ($T < 100$ K), 
with dust masses in excess of $10^8$ solar masses (Bertoldi et al. 2003,  Leipski et al. 2010).

It has been typically believed that this dust must have been produced
by core-collapse (CC) SuperNovae (SNe), as AGB stellar lifetimes
($10^8$ to $10^9$ yr) are comparable to the age of universe at
redshift $>$~6 (Morgan \& Edmunds 2003, Dwek, Galliano \& Jones 2007).
The observed mid-IR emission for a limited number of extra-galactic SNe
implies dust masses which are generally smaller than $10^{-2}~\rm
M_{\odot}$ (e.g. Sugerman et al.\ 2006, Meikle et al.\ 2007, Blair et
al. 2007, Rho et al 2008, Wesson et al. 2010a), corresponding to
condensation efficiencies which are at least two orders of magnitude
smaller than theoretical models predict (Todini \& Ferrara 2001,
Bianchi \& Schneider 2007).  FIR and sub-mm observations of dust within
supernova remnants (SNR) estimate masses ranging from $0.1-1~\rm M_{\odot}$ 
(Dunne et al. 2003, 2009, Morgan et al. 2003, Gomez et al. 2009), 
yet there are a number of difficulties with the interpretation of
these results. It is obvious that there is now indeed clear
observational evidence for dust formation in CCSNe, but the quantity
of dust {\it formed} within the ejecta is still a subject of debate.
Valiante et al. (2009) recently showed that AGB stars could
potentially rival or surpass SNe as the main producer of dust at
characteristic timescales of between 150 and 500 Myr, although the
model requires rather extreme star formation histories, a top-heavy
initial mass function and efficient condensation of dust grains in
stellar atmospheres. The dust production of SNe, either from the
progenitors (LBV, RSG, Wolf-Rayet (WR) stars) or directly in the
ejecta, versus that of AGB stars is therefore of utmost importance and
one of the science themes that will be addressed in the MESS Herschel
key program described in this paper.

\bigskip

Most of the astronomical solid state features are found in the near-IR (NIR) and
mid-IR (MIR) ranges. The {\it ISO} SWS and LWS spectrometers revolutionised our
knowledge of dust and ice around stars. In the LWS range, partly overlapping
with Herschel PACS, most of {\it ISO}s spectroscopic dust observations
suffered from signal-to-noise (S/N) problems for all but the brightest AGB stars.
The sensitivity of Herschel is a clear improvement over {\it ISO} but the short wavelength
limit of PACS ($\sim$60\,$\mu$m) is somewhat of a limitation.
Nevertheless dust-species like Forsterite (Mg$_2$SiO$_4$) at 69\,$\mu$m,
Calcite CaCO$_3$ at 92.6\,$\mu$m, Crystalline water-ice at 61\,$\mu$m, and Hibonite CaAl$_{12}$O$_{19}$
at 78\,$\mu$m are expected to be detected.  Other measured features
lack an identification e.g. the 62-63\,$\mu$m feature with
candidate substances like Dolomite, Ankerite, or Diopside (see Waters 2004 and Henning 2010 for overviews).
At longer wavelengths, PAH `drum-head' or `flopping modes' have been
predicted to occur (Joblin et al. 2002), that can be looked for with
the SPIRE FTS  (Fourier Transform Spectrometer) that will observe in an previously unexplored wavelength regime.

Apart from solid state features the PACS and SPIRE range contain a
wealth of molecular lines. Depending on chemistry and excitation
requirements, the different molecules sample the conditions in
different parts of a circumstellar envelope (CSE). 
While for example CO observations in the J= 7-6 line (370~$\mu$m) can
be obtained under good weather conditions from the ground, this line
traces gas of about 100~K. With SPIRE and PACS one can detect CO J= 45-44 
at 58.5~$\mu$m at the short wavelength edge of PACS (as was demonstrated 
in Decin et al. 2010a) which probe regions very close to the star.
Although only the Heterodyne Instrument for the Far Infrared (HIFI, de Graauw et al. 2010) onboard \it Herschel \rm will deliver resolved
spectral line observations, PACS and SPIRE with their high throughput
will allow full spectral inventories to be made.
The analysis of PACS, SPIRE (and HIFI and ground-based) molecular line
data with sophisticated radiative transfer codes (e.g. Morris et al. 1985,
Groenewegen 1994, Decin et al. 2006, 2007) will allow quantitative
statements about molecular abundances, the velocity structure in the
acceleration zone close to the star, and (variations in) the mass-loss rate.

\medskip

With these science themes in mind, the preparation for a Guaranteed
Time (GT) Key Program (KP) started in 2003, culminating in the
submission and acceptance of the MESS (Mass-loss of Evolved StarS)
GTKP in June 2007.  It involves PACS GT holders from Belgium, Austria and
Germany, the SPIRE {\it Specialist Astronomy Group} 6, and
contributions from the {\it Herschel Science Centre}, and Mission
Scientists.  The allocated time is about 300 hours, of which 170h are
devoted to imaging and the remaining to spectroscopy.

\medskip

Section~2 describes the selection of the targets and Section~3 describes the observing strategy.
Section~4 discusses some aspects of the current data reduction strategy.
Section~5 presents the key science topics that will be pursued and this is illustrated by highlights 
of the results obtained in the Science Demonstration Phase (SDP), and presenting ongoing efforts. 
Aperture photometry for 70 AGB and post-AGB stars is presented and compared to {\it AKARI} data.
Section~6 concludes this paper.
In two appendices details on the PACS mapping and data reduction strategy are presented.

\section{Target selection} 

\subsection{AGB stars and Red SuperGiants}

The main aim of the imaging program is to resolve the CSEs 
around a representative number of AGB stars, and thereby study the global
evolution of the mass-loss process and details on the structure of the CSE.
With a typical AGB lifetime of 10$^6$ year and a typical expansion
velocity of 10 \ks\ (see Habing \& Olofsson 2003) the effects of the mass-loss process could, in
principle, be traced over $3 \cdot 10^{14}$ km, or 10 pc, or about
30\arcmin\ at 1 kpc distance. In practice the outer size of the AGB
shell will be smaller, first of all due to interaction of the expanding
slow wind with the ISM, and by observational limits in terms of
sensitivity and confusion noise.

The starting point of the target selection was the {\it ISO} archive from
which all objects classified as ``stellar objects'' with SWS and LWS
observations, as well as all sources from programs which had ``AGB stars'' 
in the proposal keyword, were compiled.
In addition, stars showing extended emission in the {\it IRAS} 60 or 100 $\mu$m bands 
(Young et al. 1993) were considered as well. 
From that, a master list of about 300 objects was selected of stars seemingly
AGB stars or related to the AGB, based on the spectral type, and/or simbad classification.

The final sample was chosen to represent the various types of objects,
in terms of spectral type (covering the M-subclasses, S-stars, carbon stars), 
variability type (L, SR, Mira), and mass-loss (from low to extreme) 
within an overall allocated budget of GT for this part of the program. 
In the selection the {\it IRAS} CIRR3 flag was considered to avoid regions of high background.
Within each subclass, typically the brightest mid-IR objects were chosen.

A sample of 30 O-rich AGB stars and RSGs, 9 S-stars, and 37 C-stars will be imaged with PACS, 
as well as the two post-RSGs (IRC +10 420 and AFGL 2343).
A subset of respectively, 11, 2 and 13 AGB/RSG stars will be imaged with SPIRE, as well as R CrB, the prototype of its class
(see Table~\ref{list-all}).
That the PACS and SPIRE target lists are not identical is on the one hand a question of sensitivity--the 
fluxes are expected to be higher in the PACS wavelength domain--and on the other hand is a question of 
the available hours of guaranteed time available to the different partners.

The targets for PACS and SPIRE spectroscopy are (with one exception) a subset of the imaging targets. 
They have been selected to be bright with {\it IRAS} fluxes $S_{60} \more 50$ and $S_{100} \more 40$ Jy, a S/N 
of $\ge$ 20 on the continuum is expected over most of the wavelength range ($\sim$55 to $\sim$180 $\mu$m).
A sample of 14 O-rich AGB stars and RSG, 3 S-stars, and 6 C-stars will be observed 
spectroscopically with PACS, as well as the two post-RSGs.
A subset of respectively, 5 O-rich and 4 C-rich AGB stars and 1 post-RSG star will be observed spectroscopically with SPIRE.

\subsection{post-AGB stars and PNe}

The aim of the imaging part is very similar to that for the AGB stars: 
mapping a few infrared bright objects in the post-AGB (P-AGB) and PNe phases of
evolution, and to trace the flux and geometry of the earliest mass-loss in the phases. 
The objects proposed here are very well studied in broad wavelength
regimes, but a clear understanding of the structures of their dust
shells (and how these came about) is still lacking.

A sample of twenty well-known C- and O-rich P-AGB and PNe will be imaged with PACS, and a subset of 9 with SPIRE
(see Table~\ref{list-all}).

With respect to spectroscopic observations, the aims are again similar
to those for AGB stars, and are related mainly to molecular chemistry
and dust features.  However, another theme will be the exploitation of
the higher spatial resolution spectral mapping possibilities of PACS,
and to spatially resolve fine-structure (FS) lines of hotter P-AGB
stars and some PNe.  When the central object becomes hotter than
$T_{\rm eff}$ of around 10~000 K, FS lines become apparent 
(e.g. Liu et al. 2001, Fong et al. 2001), and even at lower effective
temperatures FS lines may arise from shocks (e.g. Hollenbach \& McKee 1989). 
Lines in the PACS domain include: [N II] 122~$\mu$m and 205~$\mu$m, 
[O III] 88~$\mu$m, [O I] 146~$\mu$m, [C II]~158 $\mu$m.  The first
three lines come from ionised regions while the fluxes of the last two
are emitted by photodissociation regions (PDRs). For the hotter PNe
(\more 20-25 kK), the FS lines from the ionised regions are good
tracers of the local conditions and abundances (e.g. Liu et al. 2001).
For instance, the [N II] 122/205 $\mu$m line-ratio gives a good tracer
for the electron density, which is not very dependent on the electron
temperature.  FS lines emitted in PDRs provide a way to determine
directly the PDR temperatures, densities, and gas masses.  The spatial
resolution of PACS may enable us to locate the ionised regions and
PDRs in the nebulae.

In total, twenty-three C- and O-rich P-AGB stars and PNe will be observed
spectroscopically with PACS, and a subset of 11 with SPIRE, of which 9
are in common with PACS.

\subsection{Massive stars}

Although massive O-type stars have intense stellar winds, their
mass-loss rates are not sufficient for them to become WR stars without
assuming short episodes of much stronger mass-loss (Fullerton et al. 2006).  
These short-lived evolutionary stages of massive star evolution 
($\sim 10^4$\,yrs for the LBV phase) produce extended regions of
stellar ejecta. In fact, most LBVs and many WR stars are nowadays
known to be surrounded by ring nebulae consisting either of ejecta or
ISM material swept up by the fast stellar winds of the
stars (e.g.  Hutsem\'ekers 1994, Nota et al. 1995, Marston 1997, Chu 2003).
These stellar winds are key to understand the stellar evolution, as
well as galactic population synthesis (e.g. Brinchmann 2010)

Most of these nebulae contain dust (Hutsem\'ekers 1997) and many of
them were detected with {\it IRAS} at colour temperatures as low as 40\,K. 
However, most of these objects were not resolved with {\it IRAS},
whereas the angular resolution of the instruments on board Herschel (6-36\arcsec)
will provide accurate maps of their far IR emission. This should help
to infer detailed information on the mass, size and structure of the
dust shells, on possible grain size / temperature gradients, therefore
significantly improving our knowledge of the mass-loss history of the
crucial LBV--WN transition phase.

PACS photometry will be obtained for several targets among the most
representative LBV (AG\,Car, HR\,Car, HD\,168625) and WR nebulae
(M\,1-67, NGC 6888). These targets represent various types of LBV and
WR nebulae at different stages of interaction with the ISM.  
PACS spectroscopy will be obtained for the brightest nebulae
to measure the fine-structure lines and determine the physical
conditions and abundances in the ionised gas and/or in the
photodissociation region.

In total, eight stars will be mapped with PACS, 2 will be observed
with PACS spectroscopy and 2 with SPIRE spectroscopy 
(see Table~\ref{list-all}, and details in Vamvatira-Nakou et al. 2011).

\subsection{Supernova Remnants}

SCUBA observations at 450 and 850~$\mu$m suggested that
0.1-1~\msol\ of cold SN dust (with a high degree of polarisation) was
present in the Cas A (Dunne et al 2003, 2009) and Kepler (Morgan et
al. 2003; Gomez et al. 2009) supernova remnants.  No cold dust
component was seen above the strong non-thermal synchrotron emission
in SCUBA observations of the Crab remnant (Green, Tuffs \& Popescu
2004).  Arguments against such large quantities of cold dust being
present in Cas A were advanced by Krause et al. (2004) and Wilson \&
Batrla (2005), where they suggested that intervening molecular clouds
along the line of sight to the remnant are responsible for the sub-mm
emission seen with SCUBA, and by Sibthorpe et al. (2010). An alternative
explanation for the excess sub-mm emission at 850~$\mu$m was proposed
by Dwek (2004), in which elongated iron needles with a high sub-mm
emissivity could also account for the excess flux without the need for
large dust masses.  We propose to address this controversial and as
yet, unresolved issue by searching for newly formed dust in five young
($<$ 1000 yr old) Galactic SNRs (this should ensure they are in the
late expansion or early Sedov phase and are therefore dominated by the
SN ejecta dynamics and not by swept up material).  We will obtain PACS
and SPIRE far-IR and sub-millimetre photometric and spectroscopic
mapping of the remnants: Cas A, Kepler, Tycho, Crab and 3C~58 (= SN 1181); 
the spectroscopic insight will be important in helping to disentangle
the issue of interstellar versus supernova and/or stellar wind material.


\begin{table*}[h]
\setlength{\tabcolsep}{1.0mm}
\footnotesize
\caption{Target List. The spectral type comes from the SIMBAD database. Sources are listed alphabetically per sub-class.}
\label{list-all}
\begin{tabular}{llrrrrrc}
\hline
IRAS name    & Identifier                       & type        & PACS    & PACS  & SPIRE  & SPIRE  & Remarks   \\
             &                                  &             & Imag.    & Spec. & Imag.  & Spec.  \\ \hline

O-rich AGB \\ \hline
01556+4511   &  AFGL 278                       &       M7III  & $\surd$ &   &   &   & \\
17411-3154   &  AFGL 5379                      &    M         &   &  $\surd$ &    &   & \\
05524+0723   &  $\alpha$ ORI / Betelgeuse      &      M1Iab:  & $\surd$ &  $\surd$ &  $\surd$ &  $\surd$ & \\
21439-0226   &  EP AQR                         &    M8IIIvar  & $\surd$ &   &   &   & \\
20077-0625   &  IRC -10 529 / AFGL 2514        &          M:  & $\surd$ &  $\surd$ &    &   & \\
21419+5832   &  $\mu$ CEP                      &       M2Iae  & $\surd$ &    & $\surd$    &   & \\
             &  NML CYG                        &      M6IIIe  & $\surd$ &  $\surd$ &    &   & \\
03507+1115   &  NML TAU / IK Tau               &        M6me  & $\surd$ &  $\surd$ &  $\surd$ &  $\surd$ & \\
02168-0312   &  $o$ CET                        &      M7IIIe  & $\surd$ &  $\surd$ &  $\surd$ &  $\surd$ & \\
01304+6211   &  OH 127.8+0.0                   &           M  & $\surd$ &   &   &   & \\
18348-0526   &  OH 26.5+0.6                    &           M  & $\surd$ &  $\surd$ &    &   & \\
23412-1533   &  R AQR                          &  M7IIIpevar  & $\surd$ &    &  $\surd$ &   & \\
23558+5106   &  R CAS                          &      M7IIIe  & $\surd$ &  $\surd$ &  $\surd$ &   & \\
10580-1803   &  R CRT                          &       M7III  & $\surd$ &   &   &   & \\
04361-6210   &  R DOR                          &      M8IIIe  & $\surd$ &  $\surd$ &  $\surd$ &   & \\
13269-2301   &  R HYA                          &      M7IIIe  & $\surd$ &    &  $\surd$ &   & \\
09448+1139   &  R LEO                          &      M8IIIe  & $\surd$ &    &    &   & \\
13001+0527   &  RT VIR                         &       M8III  & $\surd$ &   &   &   & \\
14219+2555   &  RX BOO                         &        M7.5  & $\surd$ &    &  $\surd$ &   & \\
13114-0232   &  SW VIR                         &       M7III  & $\surd$ &   &   &   & \\
14003-7633   &  $\theta$ APS                   &    M6.5III:  & $\surd$ &   & $\surd$   &   & \\
20248-2825   &  T MIC                          &       M7III  & $\surd$ &   &   &   & \\
04566+5606   &  TX CAM                         &        M8.5  & $\surd$ &  $\surd$ &    &   & \\
20038-2722   &  V1943 SGR                      &       M7III  & $\surd$ &   &   &   & \\
04020-1551   &  V ERI                          &     M5/M6IV  & $\surd$ &   &   &   & \\
18050-2213   &  VX SGR                         &   M5/M6III:  & $\surd$ &    &    &   & \\
07209-2540   &  VY CMA                         &    M3/M4II:  & $\surd$ &  $\surd$ &    &  $\surd$ & a  \\
13462-2807   &  W HYA                          &         M7e  & $\surd$ &  $\surd$ &  $\surd$ &  $\surd$ & \\
01037+1219   &  WX PSC / IRC +10 011           &         M9:  & $\surd$ &  $\surd$ &    &   & \\
16011+4722   &  X HER                          &          M8  & $\surd$ &  $\surd$ &   &   & \\
20075-6005   &  X PAV                          &  M6/M7III:p  & $\surd$ &   &   &   & \\
             &                                 &              &   &   &   &   & \\
S-stars AGB \\ \hline
19486+3247   &  $\chi$ CYG                     &           S  & $\surd$ & $\surd$ &   &   & \\
04497+1410   &  $o$ ORI                        &        S+WD  & $\surd$ &   &   &   & \\
22196-4612   &  $\pi$ GRU                      &      S5+G0V  & $\surd$ & $\surd$ & $\surd$ &   & \\
09076+3110   &  RS CNC                         &      M6IIIS  & $\surd$ &   & $\surd$ &   & \\
20120-4433   &  RZ SGR                         &        S4,4  & $\surd$ &   &   &   & \\
01159+7220   &  S CAS                          &          Se  & $\surd$ &   &   &   & \\
15492+4837   &  ST HER                         &         M6s  & $\surd$ &   &   &   & \\
19126-0708   &  W AQL                          &     S6.6+FV  & $\surd$ & $\surd$ &   &   & \\
07245+4605   &  Y LYN                          &    M5Ib-IIS  & $\surd$ &   &   &   & \\

\hline
\end{tabular}
\end{table*}

\setcounter{table}{0}
\begin{table*}[h]
\setlength{\tabcolsep}{1.0mm}
\footnotesize
\caption{Continued}
\label{list-all}
\begin{tabular}{llrrrrrc}
\hline
IRAS name    & Identifier                      & type        & PACS     & PACS       & SPIRE   & SPIRE  & Remarks \\
             &                                 &             & Imag.    & Spec.      & Imag.   & Spec.    \\ \hline

C-rich AGB \\ \hline
01144+6658   &  AFGL 190                       &           C  & $\surd$ &   &   &   & \\
18240+2326   &  AFGL 2155                      &           C  & $\surd$ &   &   &   & \\
23166+1655   &  AFGL 3068                      &           C  & $\surd$ & $\surd$ &  $\surd$ &  $\surd$ & \\
23320+4316   &  AFGL 3116 / IRC +40 540        &           C  & $\surd$ & $\surd$ &   &   & \\
00248+3518   &  AQ AND                         &        Nvar  & $\surd$ &   & $\surd$ &   & \\
19314-1629   &  AQ SGR                         &         CII  & $\surd$ &   &   &   & \\
10131+3049   &  CIT 6                          &           C  & $\surd$ & $\surd$ &  $\surd$ &  $\surd$ & \\
09452+1330   &  CW LEO / IRC +10 216           &           C  & $\surd$ & $\surd$ &  $\surd$ &  $\surd$ & a, b  \\
11331-1418   &  HD 100764                      &        C-R.  & $\surd$ &   &   &   & \\
15194-5115   &  IRAS 15194-5115                &           C  & $\surd$ & $\surd$ &   &   & \\
02270-2619   &  R FOR                          &           C  & $\surd$ &   &   &   & \\
04573-1452   &  R LEP                          &        CIIe  & $\surd$ &   &   &   & \\
01246-3248   &  R SCL                          &         CII  & $\surd$ &   &  $\surd$ & $\surd$ & \\
20141-2128   &  RT CAP                         &         CII  & $\surd$ &   &   &   & \\
12544+6615   &  RY DRA                         &           C  & $\surd$ &   &   &   & \\
21358+7823   &  S CEP                          &         CII  & $\surd$ &   & $\surd$ &   & c  \\
18476-0758   &  S SCT                          &         CII  & $\surd$ &   &   &   & \\
04459+6804   &  ST CAM                         &           N  & $\surd$ &   &   &   & \\
18306+3657   &  T LYR                          &           C  & $\surd$ &   &   &   & \\
19390+3229   &  TT CYG                         &           N  & $\surd$ &   & $\surd$ &   & c \\
03112-5730   &  TW HOR                         &         CII  & $\surd$ &   &   &   & \\
23438+0312   &  TX PSC                         &         CII  & $\surd$ &   & $\surd$ &   & \\
10329-3918   &  U ANT                          &       N:var  & $\surd$ &   &  $\surd$ &   & \\
03374+6229   &  U CAM                          &           N  & $\surd$ &   & $\surd$ &   & \\
10350-1307   &  U HYA                          &         CII  & $\surd$ &   &   &   & \\
06331+3829   &  UU AUR                         &         CII  & $\surd$ &   &   &   & \\
19233+7627   &  UX DRA                         &         CII  & $\surd$ &   & $\surd$ &   & \\
15477+3943   &  V CRB                          &           N  & $\surd$ &   &   &   & \\
20396+4757   &  V CYG                          &           N  & $\surd$ & $\surd$ &   &   & \\
10491-2059   &  V HYA                          &           C  & $\surd$ & $\surd$ &  $\surd$ &   & \\
17389-5742   &  V PAV                          &          C+  & $\surd$ &   &   &   & \\
10416+6740   &  VY UMA                         &         CII  & $\surd$ &   &   &   & \\
05028+0106   &  W ORI                          &         CII  & $\surd$ &   &   &   & \\
05418-4628   &  W PIC                          &        Nvar  & $\surd$ &   &   &   & \\
15094-6953   &  X TRA                          &           C  & $\surd$ &   &   &   & \\
12427+4542   &  Y CVN                          &       CIab:  & $\surd$ &   &  $\surd$ &   & \\
21197-6956   &  Y PAV                          &         CII  & $\surd$ &   &   &   & \\
             &                                 &              &   &   &   &   & \\
Post-RSG \\ \hline
19114+0002   &  HD 179821 / AFGL 2343          &        G5Ia  & $\surd$ & $\surd$ &   & $\surd$  &  \\
19244+1115   &  IRC +10 420                    &        F8Ia  & $\surd$ & $\surd$ &   &   &  \\
             &                                 &              &   &   &   &   & \\
R CrB \\ \hline
15465+2818   &  R CRB                          &    G0Iab:pe  &   &   & $\surd$ &   & \\

\hline
\end{tabular}
\end{table*}

\setcounter{table}{0}
\begin{table*}[h]
\setlength{\tabcolsep}{1.0mm}
\footnotesize
\caption{Continued}
\label{list-all}
\begin{tabular}{llrrrrrc}
\hline
IRAS name    & Identifier                      & type        & PACS & PACS & SPIRE   & SPIRE  & Remarks \\
             &                                 &             & Imag. & Spec.& Imag.   & Spec.    \\ \hline
Post-AGB \\ \hline
             &  AFGL 2688                      &  C-PAGB      & $\surd$ & $\surd$ & $\surd$ &  $\surd$ & \\
10215-5916   &  AFGL 4106                      &    PAGB      &   &   &   &  $\surd$ & \\
04395+3601   &  AFGL 618                       &  C-PAGB      & $\surd$ & $\surd$ & $\surd$ &  $\surd$ & \\
17150-3224   &  AFGL 6815                      &      O-PAGB  & $\surd$ & $\surd$ &   &   & \\
09371+1212   &  Frosty Leo Neb.                &  PAGB        &         & $\surd$ &   &  $\surd$ & \\
11385-5517   &  HD 101584                      &  PAGB        & $\surd$ &  &   &    & \\
17436+5003   &  HD 161796                      &  O-PAGB      & $\surd$ & $\surd$ & $\surd$ &  $\surd$ & \\
22272+5435   &  HD 235858                      &  C-PAGB      & $\surd$ & $\surd$ &   &   &  \\
06176-1036   &  HD 44179 / Red Rectangle       &  C-PAGB      & $\surd$ & $\surd$ & $\surd$ &  $\surd$ & b \\
07134+1005   &  HD 56126                       &  C-PAGB      & $\surd$ & $\surd$ & $\surd$ &  $\surd$ &  \\
13428-6232   &  IRAS 13428-6232                &     PAGB     & $\surd$ & $\surd$ &   &   & \\
16342-3814   &  IRAS 16342-3814                &      O-PAGB  & $\surd$ & $\surd$ &   &   & \\
16594-4656   &  IRAS 16594-4656                &     C-PAGB   & $\surd$ &   &   &   & \\
19343+2926   &  M 1-92                         &  PAGB        & $\surd$ &  &   &    & \\
07399-1435   &  OH 231.8+4.2                   &      O-PAGB  & $\surd$ & $\surd$ &   &  $\surd$ & \\
          &                                 &              &   &   &   &   & \\
PNe  \\ \hline
19327+3024   &  BD +30 3639                    &  PN          &   & $\surd$ &   &   & \\
17047-5650   &  CPD -56 8032                   &    PN        &   &   &   &  $\surd$ &  \\
17347-3139   &  GLMP 591                       &  PN          &   & $\surd$ &   &   & \\
14562-5406   &  Hen 2-113                      &  PN/WC       &   & $\surd$ &   &   & \\
17423-1755   &  Hen 3-1475                     &  PAGB/Be     &   & $\surd$ &   &   & \\
10178-5958   &  Hen 3-401                      &  PN/Be       &   & $\surd$ &   &   &  \\
09425-6040   &  IRAS 09425-6040                &  PAGB        &   & $\surd$ &   &   & \\
16279-4757   &  IRAS 16279-4757                &  PAGB        &   & $\surd$ &   &   & \\
22036+5306   &  IRAS 22036+5306                &  PN          &  $\surd$  & $\surd$ &   &   & \\
11119+5517   &  NGC 3587                       & C-PN         & $\surd$ &   & $\surd$ &    & \\
17103-3702   &  NGC 6302                       &     O-PNe    &         & $\surd$ &   &  $\surd$ &  \\
01391+5119   &  NGC 650                        & PN           & $\surd$ &   & $\surd$ &    & \\
18021-1950   &  NGC 6537                       &      O-PNe   &         & $\surd$ &   &    & \\
17584+6638   &  NGC 6543                       &      PN      & $\surd$ &   & $\surd$ &    & \\
18517+3257   &  NGC 6720 / Ring Nebula         &       O-PNe  & $\surd$ &   & $\surd$ &    & b  \\
19574+2234   &  NGC 6853                       & PN           & $\surd$ &   & $\surd$ &    & \\
             &  NGC 7027                       &      C-PNe   & $\surd$ & $\surd$ &   &  $\surd$ & a  \\
22267-2102   &  NGC 7293 / Helix Nebula        & PN           & $\surd$ &   & $\surd$ &    & e \\
21282+5050   &  PN G093.9-00.1                 &  PN          &   & $\surd$ &   &   & \\
             &                                 &              &   &   &   &   & \\
Massive stars \\ \hline
10541-6011   &  AG Car                         &   LBV        & $\surd$ & $\surd$ &   & $\surd$ & \\
             &  G79.29+0.46                    &   LBV?       & $\surd$ &   &   &   & \\
18184-1623   &  HD 168625                      &   LBV?       & $\surd$ & $\surd$ &   &   & \\
10520-6010   &  He 3-519                       &   Of/WN      & $\surd$ &   &   &   & \\
10211-5922   &  HR Car                         &   LBV        & $\surd$ &   &   &   & \\
19092+1646   &  M 1-67                         &   WN8        & $\surd$ &   & $\surd$ & $\surd$ & \\
             &  NGC 6888                       &   WN6        & $\surd$ &   &   &   & \\
11065-6026   &  Wra 751                        &   LBV        & $\surd$ &   &   &   & \\
             &                                 &              &   &   &   &   & \\
SNe remnants  \\ \hline
             &  Cas A / SNR 111.7-02.1         &              & $\surd$ & $\surd$ & $\surd$ & $\surd$ & d \\
05314+2200   &  Crab / SNR 184.6-05.8          &              & $\surd$ & $\surd$ & $\surd$ & $\surd$ & \\
17276-2126   &  Kepler's SN / SNR 004.5+06.8   &              & $\surd$ & $\surd$ & $\surd$ &  & \\
             &  SN 1181 / SNR 130.7+03.1       &              & $\surd$ & $\surd$ &  &  & \\
             &  Tycho's SN / SNR 120.1+01.4    &              & $\surd$ & $\surd$ & $\surd$ & $\surd$ & \\

\hline
\end{tabular}

Remarks:
(a) PACS+SPIRE spectroscopy obtained as part of PV.
(b) PACS+SPIRE imaging obtained as part of SDP (data are public).
(c) PACS imaging obtained as part of SDP (data are public).
(d) PACS+SPIRE imaging obtained as part of SDP.
(e) Data taken in PACS-SPIRE parallel mode. \\

\end{table*}

\section{Observing strategy}

\subsection{SPIRE imaging} 

Each of the 27 evolved star targets will be mapped in ``Large Map'' mode,
with a scan leg length of 30\arcmin, a cross-scan length of 30\arcmin\
and a repetition factor of 3. 
Such an observation takes 4570 sec of which 2800 sec are spent on-source.
The 5 SNe remnants will be mapped in the same 
mode and repetition factor over 32\arcmin$\times$32\arcmin\ to ensure sufficient sky coverage.
This should yield 5$\sigma$ detections
for a point source with fluxes of 26, 22 and 31 mJy at 250, 350 and
500 $\mu$m, or 5$\sigma$ per beam for an extended source with surface
brightnesses of 6.0, 4.5 and 2.0 MJy~sr$^{-1}$ at these wavelengths, and will reach the confusion noise at these wavelengths.
These numbers, and the sensitivities quoted in the next three
  sub-sections, are based on HSPOT \it Herschel \rm Observation Planning Tool v5.1.1, which includes calibration
  files based on the in-flight performance of the instruments.  

One object (the Helix Nebula) is so large that it will be observed in
SPIRE-PACS parallel mode over an area of about 75\arcmin\ squared.

As some imaging data obtained in SDP was made public, 
a few hours of observing time could be put back into the program as compensation. 
Based on the PACS observations available in May 2010, it was decided
to obtain a 8\arcmin$\times$8\arcmin\ SPIRE map of M1-67 (a massive  object), 
a 15\arcmin$\times$15\arcmin\ map of NGC 6853 and
10\arcmin$\times$10\arcmin\ maps of NGC 650 and 8 AGB targets.

The flux calibration uncertainty is 15\% for all SPIRE photometer bands 
at the time these data were reduced (Swinyard et al. 2010).  The three 
SPIRE beams are approximately Gaussian with FWHM sizes of 18.1, 25.2, 
and 36.6\arcsec, and an uncertainty of 5\%.

\subsection{SPIRE spectroscopy} 

A single pointing FTS observation will be obtained for each of
the 23 evolved star targets, with sparse image sampling. The high+low spectral
resolution setting (the highest unapodised spectral resolution is 0.04 cm$^{-1}$) will be used, with a repetition factor of 17 for each mode.

For a source with fluxes of 24, 5 and 1.8 Jy at 200, 400 and
600~$\mu$m, this is predicted to yield continuum S/N ratios per
resolution element of 57, 18 and 4.8, respectively, in the
high spectral resolution mode.
Such an observation takes 2710 sec in total, of which 2260 sec are spent on-source.

For the Cas~A supernova remnant, FTS spectroscopy in high resolution mode 
will be obtained at three positions on the remnant, each sparsely sampling 
the 2.6\arcmin\ field of view of the FTS, with a repetition factor of 24. 
Similar spectra will be obtained for one pointing position only for 
each of the Crab and Tycho supernova remnants.

\subsection{PACS imaging} 

All PACS imaging is done using the ``scan map'' Astronomical Observation Request (AOR) 
with the ``medium'' scan speed of 20\arcsec/s (with the exception of the Helix nebulae mentioned 
earlier that will be observed using PACS-SPIRE parallel mode at 60\arcsec/s speed).
Our observations are always the concatenation of 2 AORs (a scan and an orthogonal cross-scan).

The P-AGB and PNe objects will be imaged at 70 and 160 $\mu$m.
The size of the maps range from about 4 to 9 arcmin on a side with a repetition factor of 6-8.
The five SNe remnants will be observed with 22\arcmin\ scan-leg
lengths at 70, 100 and 160$~\mu$m with an repetition factor of 2.

For the AGB stars, the largest sub-sample within the program, dust
radiative transfer calculations have been performed for all sources
with the code DUSTY (Ivezi\'c et al. 1999), to predict the total flux
and the surface brightness as a function of radial distance to the
star at the PACS and SPIRE wavelengths under the assumption of a
constant mass-loss rate, taking literature values for the stellar
parameters, like effective temperature and distance. The current mass-loss
rate and the grain properties were determined by fitting the SED,
using archival {\it ISO} SWS + LWS spectra (or {\it IRAS} LRS spectra if no {\it ISO}
spectra were available) and broad-band photometry from the optical and
infra red and including sub-mm fluxes when available, as constraints. 
This work is described in detail in the Ph.D. thesis by D.~Ladjal (2010).
The model predictions were then compared to the background and
confusion noise estimates coming from HSPOT as well as by inspecting the
IRAS Sky Survey Atlas\footnote{http://irsa.ipac.caltech.edu/applications/IRAS/ISSA/} 
(Wheelock et al. 1994) 100 $\mu$m maps to arrive at an optimal map size.
In the end, the source size was determined by the radius where a 1$\sigma$ noise at 70 $\mu$m of 3 mJy/beam was predicted.
Before submission of the final AORs, we also took advantage of the
fact that images obtained with {\it Spitzer} and {\it AKARI} revealed 
interaction of the CSE with the ISM for some targets 
(Ueta et al. 2008, 2010, Izumiura et al. 2009) and these set the minimum size of the maps.

The AGB objects will be imaged at 70 and 160 $\mu$m and the scan-length varies between 6 and 34\arcmin.
Repetition factors between 2 and 8 are used. For reference, a single scan over a typical area with a 
repetition factor of 1 results in a 1-$\sigma$ point-source sensitivity of 12, 14, 26 mJy 
or 11.9, 11.5, 5.3 MJy~sr$^{-1}$ in the 70, 100 and 160~$\mu$m bands, and this is expected to decrease 
with 1/$\sqrt{n}$ for the number of repetitions considered here. As a
scan and cross-scan are always concatenated there is another factor $\sqrt{2}$ gain in sensitivity.
The duration of a concatenated scan plus cross-scan for a typical
  observation with a scan-leg of 15\arcmin, a cross-scan step of
  155\arcsec\ (see Appendix~A for details) and a repetition factor of
  3 takes 2420 sec of which 1350 sec are spent on-source.  Every point
  in the sky located in the central part of the map that is covered
  approximately in a homogeneous way (see Appendix~A for details) is
  observed theoretically for about 31 sec. 
The typical confusion noise in the three PACS bands are 0.2-0.4, 0.5-1.2, 0.8-1.8 MJy~sr$^{-1}$, which implies that 
the deepest maps we will carry out may just reach the confusion noise in the reddest band, but never at 
shorter wavelengths, contrary to SPIRE where the confusion limit is reached in all bands after a 
few ($\more$ 3) repetitions (Nguyen et al. 2010).

The flux calibration uncertainty currently is 10\% for the 70 and 100
$\mu$m bands and better than 15\% at 160 $\mu$m (Poglitsch et
al. 2010).  The PACS beams are approximately Gaussian with FWHM sizes
of 5.6, 6.8, and 11.4\arcsec, at 70, 100 and 160 $\mu$m, respectively
(at a scanspeed of 20\arcsec/s).  
At a low flux level a tri-lobe structure in the PSF is seen (see
  Fig~\ref{Fig-aqand}), that is explained by the secondary mirror
  support structure, see Pilbratt et al. (2010) and Poglitsch et
  al. (2010).

\subsection{PACS spectroscopy} 

All PACS spectroscopy is performed by concatenating one (or more) 
(B2A+short R1) and (B2B+long R1) AORs to cover the full PACS wavelength range 
from about 54 to 210 $\mu$m with Nyquist wavelength sampling.
The duration of these 2 concatenated AORs is 3710 sec of which 3130 sec are spent on-source.
The observing mode is pointed with chopping/nodding with a medium chopper 
throw of 3\arcmin.

The sensitivity of the PACS spectrometer is a complicated function of wavelength 
(see Section 4 in the PACS User Manual, which is available through the {\it Herschel Science Centre}) 
but is best in the 110-140 $\mu$m range at about 0.8~Jy (1-$\sigma$) continuum sensitivity for 
a repetition factor of one.
The spectral resolution also strongly depends on wavelength (see Section 4 in the PACS User Manual) 
and is lowest at about $R$= 1100 near 110 $\mu$m and highest at about $R$= 5000 near 70 $\mu$m.

Four sources were submitted as SDP targets for PACS spectroscopy, but 
three (NGC 7027, VY CMa and CW Leo) happened to be observed already as 
part of the Performance Verification (PV) of the AORs. These data were officially ``transferred'' to 
the MESS program. Two of these sources (VY CMa and CW Leo) were observed 
in mapping mode. All details over these maps can be found in Royer et al. 
(2010) and Decin et al. (2010b) respectively.

Nine positions on the Cas~A supernova remnant will be observed with
the PACS IFU, obtaining full spectral coverage using the 
(B2A+short R1) and (B2B+long R1) AORs and the largest chopper throw of 6\arcmin. 
Single positions on each of the Crab, Kepler and Tycho remnants will be observed in the same way.

\section{Data reduction}

\subsection{PACS imaging} 
\label{pacs-ima}

Poglitsch et al. (2010) describe the standard data reduction scheme for 
scan maps from so-called ``Level 0'' (raw data) to ``Level 2'' products.  
Level 1 and 2 products that are part of the data sets retrieved from
the {\it Herschel} Science Archive (HSA) have been produced by
execution of the standard pipeline steps, as described by Poglitsch et al. (2010)

We do not use the products generated in this way, but use an
adapted and extended version of the pipeline script suited to our needs starting from the raw data. 
In particular the deglitching step(s), and the ``high pass filtering''
need special care. Therefore, the making of the map has become a
two-step process. In addition, as the standard pipeline script
operates on a single AOR, while our observations are always the
concatenation of 2 AORs (a scan and an orthogonal cross-scan) an additional step
is also needed.  Details of the reduction steps can be found in Appendix~B.


\subsection{PACS spectroscopy} 

Poglitsch et al. (2010) describe the standard data reduction scheme for 
chopped spectroscopic measurements and the spectroscopic results described in 
Sect.~\ref{S-science} have been obtained using the nominal pipeline
except that we combined the spectra obtained at the two nod positions
after rebinning only (Royer et al. 2010).

\subsection{SPIRE imaging} 

The standard SPIRE photometer data processing pipeline, described by
Griffin et al. (2008, 2010) and Dowell et al. (2010), is sufficient to reduce our SPIRE
photometric imaging data. The calibration steps for these data are
described by Swinyard et al. (2010). The high stability and low 1/$f$
noise knee frequency of the SPIRE detectors mean that no filtering of
the signal time-lines is required.  The larger beam size and
characteristic response of the SPIRE bolometers to cosmic rays means
that the standard deglitching routine is applicable in this case.

The only custom step required in the reduction of these data is the 
removal of the bolometer base-lines.  The nominal processing removes 
only the median from each time-line.  In the presence of large scale 
structure and/or galactic cirrus, this method can result in striping due 
to sub-optimal baseline estimation.  To overcome this we implement the 
iterative baseline removal described by Bendo et al. (2010).

This method starts by creating an initial map from the median subtracted 
signal time-lines.  The signal in a given bolometer time-line is then 
compared to the value of the derived map pixel for each pointing within 
a scan-leg.  The median difference is then calculated, and subtracted 
from the time-line.  This process is repeated 40 times to fully remove 
the striping.  Finally the median is removed from the last iteration to 
give the final map product.

\subsection{SPIRE spectroscopy} 

Griffin et al. (2010) and Fulton et al. (2010) describe the SPIRE
spectrometer data processing pipeline, while Swinyard et al. (2010)
describe how the spectra are calibrated.  We briefly review here the data
reduction of the MESS targets.

Data reduction was carried out in HIPE (\it Herschel \rm Interactrive Processing Environment, Ott et al. 2010).  
An important step in SPIRE spectral data reduction is accurate
subtraction of the background emission from the telescope and
instrument.  During Herschel's SDP, a reference interferogram was
obtained on the same Operational Day (OD) as target observations.  The
sources observed at these times were also relatively bright. The
subtraction of a reference interferogram obtained close in time to the
target spectrum gave good results.

Since the beginning of routine observations, fainter targets have been
observed, and deep reference interferograms have been obtained less
frequently. If the temperature of the telescope and instrument differ
significantly between the dark observation and the source observation,
unwanted spectral artefacts can appear, the most obvious being a
ski-slope-like shape to the continuum in the SLW band.

Routine phase calibrations now use a "superdark" created by combining
a large number of dark observations to obtain a very high
signal-to-noise ratio.  A short dark observation obtained on the same
OD as the source spectrum is then used to scale the superdark to 
correct for the variations in instrument and telescope temperatures.

Data in the SDP were flux-calibrated using observations of the
asteroid Vesta, due to the inaccessibility at that time of Uranus and
Neptune, the preferred calibration sources.  Routine phase flux
calibration uses observations of Uranus and the planetary atmosphere
models of Moreno (1998, 2010).  The Vesta-based overall calibration
accuracy was estimated at 15-30\% (Griffin et al. 2010); the
Uranus-based calibration is a significant improvement, particularly in
the SLW band.

The FTS produces spectra in which the line profiles are sinc
functions; to remove the negative side-lobes of the sinc functions, we
apodize our spectra using the extended Norton-Beer 1.5 apodizing
function described in Naylor \& Tahic (2007).  This gives line
profiles which are close to Gaussian, with a full width at
half-maximum (FWHM) of 0.070~cm$^{-1}$ (2.1 GHz).

\section{Key science and first results} 
\label{S-science}

The results obtained during SDP and the early phases of Routine Phase (RP) match and exceed our expectations.
Eight papers were published in the A\&A special issue and one in {\it Nature} 
that illustrate well the science that will be pursued.

\subsection{Supernova dust}

Barlow et al. (2010) analyse PACS and SPIRE images of the CC~SN
remnant Cas~A to resolve for the first time a cool dust component,
with an estimated mass of 0.075~\msol, significantly lower than
previously estimated for this remnant from ground-based sub-millimeter
observations (see Section 2.4) but higher than the dust masses derived
for nearby extragalactic supernovae from {\em Spitzer} observations at
mid-IR wavelengths (see Section~1).

\subsection{The circumstellar envelopes}

Van Hoof et al. (2010) analyse PACS and SPIRE images of the planetary nebula NGC 6720 (the Ring nebula). 
There is a striking resemblance between the dust distribution and H$_2$ emission (from ground based data), 
which appears to be observational evidence that H$_2$ has been formed on grain surfaces. 
They conclude that the most plausible scenario is that the H$_2$ resides in high density knots 
which were formed after recombination of the gas started when the central star entered the cooling track.

Another main result is on the detached shells around AGB stars that
were first revealed by the {\it IRAS} satellite as objects that showed
an excess at 60 $\mu$m. Later observations by single-dish (Olofsson et al. 1996) 
and interferometric millimetre telescopes (Lindqvist et al. 1999, Olofsson et al. 2000) 
revealed spherical and thin shells emitting in CO (TT Cyg, U Ant, S Sct, R Scl, U Cam), that 
were interpreted as short phases  of high mass loss, probably related to thermal pulses.
Kerschbaum et al. (2010) present for the first time resolved (PACS) images 
of the dust shells around AQ And, U Ant, and TT Cyg, which allows the
derivation of the dust temperature at the inner radius.  Although U Ant 
and TT Cyg are classical examples of stars surrounded by a detached shell, 
AQ And was not previously known to have such a shell. 
Additional examples are presented in Kerschbaum et al. (2011) and Mecina et al. (2011).

\subsection{Wind-ISM interaction}

\begin{figure} 

\begin{minipage}{0.49\textwidth}
\resizebox{\hsize}{!}{\includegraphics{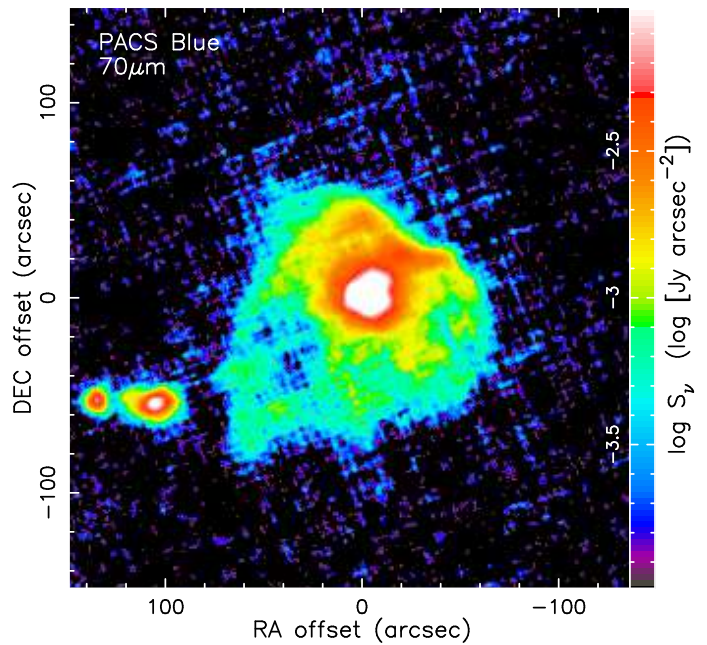}}
\end{minipage}


\caption[]{ 
The map of X Her at 70 $\mu$m displayed on a logarithmic intensity scale.
North is to the top, east to the left.
To the south east of X Her are a pair of galaxies (UGC10156a and b).
} 
\label{Fig-xher} 
\end{figure}

Sahai \& Chronopoulos (2010) present ultraviolet {\it GALEX} images of
IRC +10 216 (CW Leo) that revealed for the first time a bow shock.
Ladjal et al. (2010) demonstrate that the bow shock is also visible in
PACS and SPIRE images. The dust associated with the bow shock has a
temperature estimated at 25 K. Using the shape of the shock and a
published proper motion, a space motion of CW Leo relative to the ISM 
of $107 /\sqrt{n_{\rm ISM}}$ \ks\ is derived.  A comparison to the models 
by Wareing et al. (2007) regarding the shape  of the bow shock suggests 
that the space motion is likely to be \less 75 \ks, implying a local ISM density \more 2 cm$^{-3}$.
In fact, examples of bow shocks are quite common in the MESS sample.
Figure~\ref{Fig-xher} shows the case of X Her (Jorissen et al., in prep.) 
which displays extended emission over about 2\arcmin. Young et
  al. (1993) quote an extension in the IRAS 60 $\mu$m band of more
  than 12\arcmin\ but we find no evidence for that.  Interestingly,
  panel B in their Figure~11 shows an intensity profile for X Her which
  appears to be much narrower than the size quoted in their Table~3.
  Additionally, IRAS did not have the spatial resolution to resolve
  the pair of interacting galaxies we find, possibly leading to an
  overestimate of the size of the extended emission by Young et al.
In the case of X Her the flattened shape of the bow shock suggests a
larger space motion and/or a larger ISM density compared to CW Leo, or
to a highly inclined orientation of the bow shock surface with respect
to observers.  Additional examples are presented in Jorissen et
al. (2011) and Mayer et al. (2011).

\subsection{Molecular lines and chemistry}

Regarding spectroscopy, the results published so far illustrate 
the enormous potential to detect new molecular lines thanks to the
increased spectral resolution of PACS w.r.t. {\it ISO}s LWS and the
previously unexplored wavelength region covered by the SPIRE FTS.

Three papers concentrate on CW Leo. 
Cernicharo et al. (2010) discuss the detection of HCl in this object,
and derive an abundance relative to H$_2$ of 5$\times$10$^{-8}$ and
conclude that HCl is produced in the innermost layers of the
circumstellar envelope and extends until the molecule is
photodissociated by interstellar UV radiation at about 4.5 stellar radii.
Decin et al. (2010a) discover tens of lines of SiS and SiO including
very high transitions that trace the dust formation zone.  A comparison 
to chemical thermo dynamical equilibrium models puts constrains on the 
fraction of SiS and SiO that are involved in the dust formation process.
The discovery of water a few years ago by Melnick et al. (2001)
spurred a lot of interest. However the exact origin could so far not
be identified. Decin et al. (2010b) analyse SPIRE and PACS
spectroscopy and present tens of water lines of both low and high
excitation.  A possible explanation is the penetration of
interstellar ultraviolet photons deep into a clumpy CSE initiating an
active photo chemistry in the inner envelope.

\begin{figure} 

\begin{minipage}{0.49\textwidth}
\resizebox{\hsize}{!}{\includegraphics{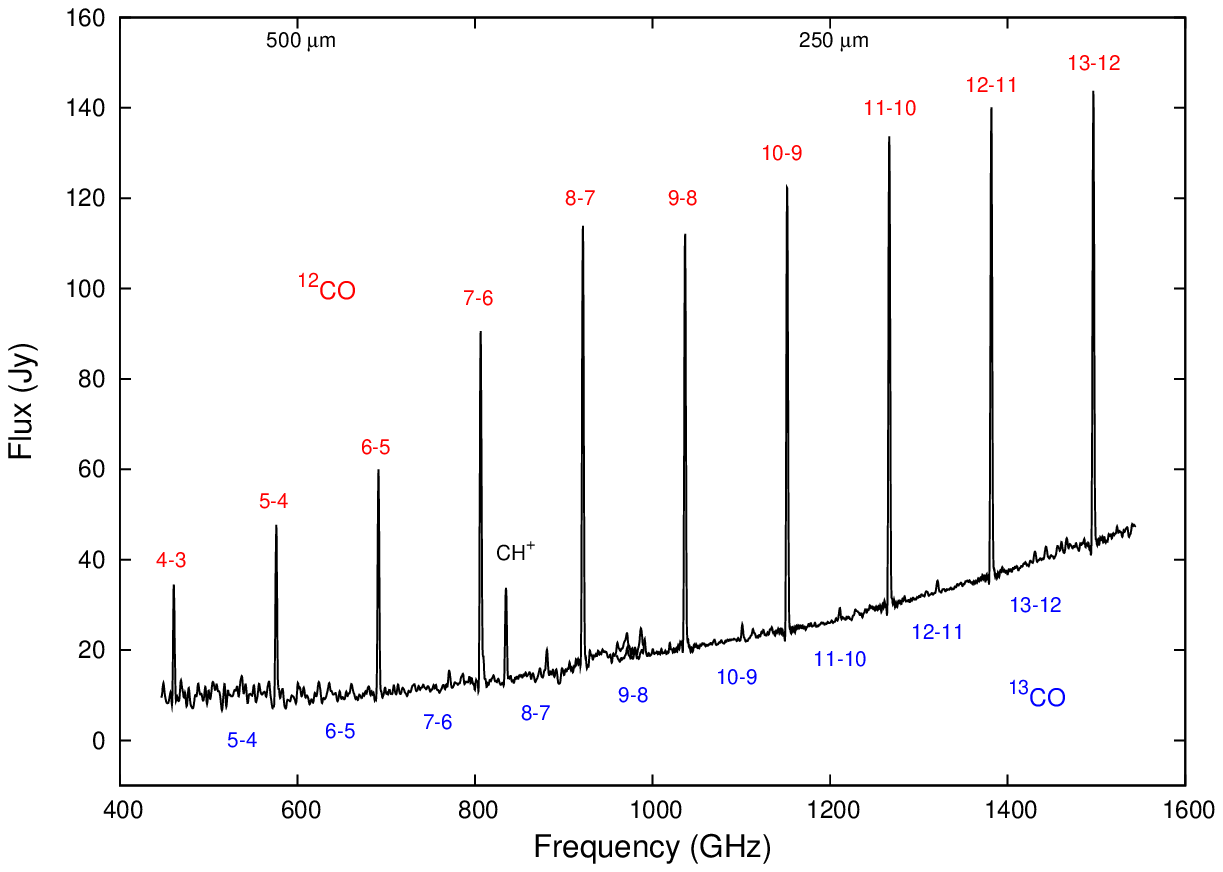}}
\end{minipage}

\begin{minipage}{0.49\textwidth}
\resizebox{\hsize}{!}{\includegraphics{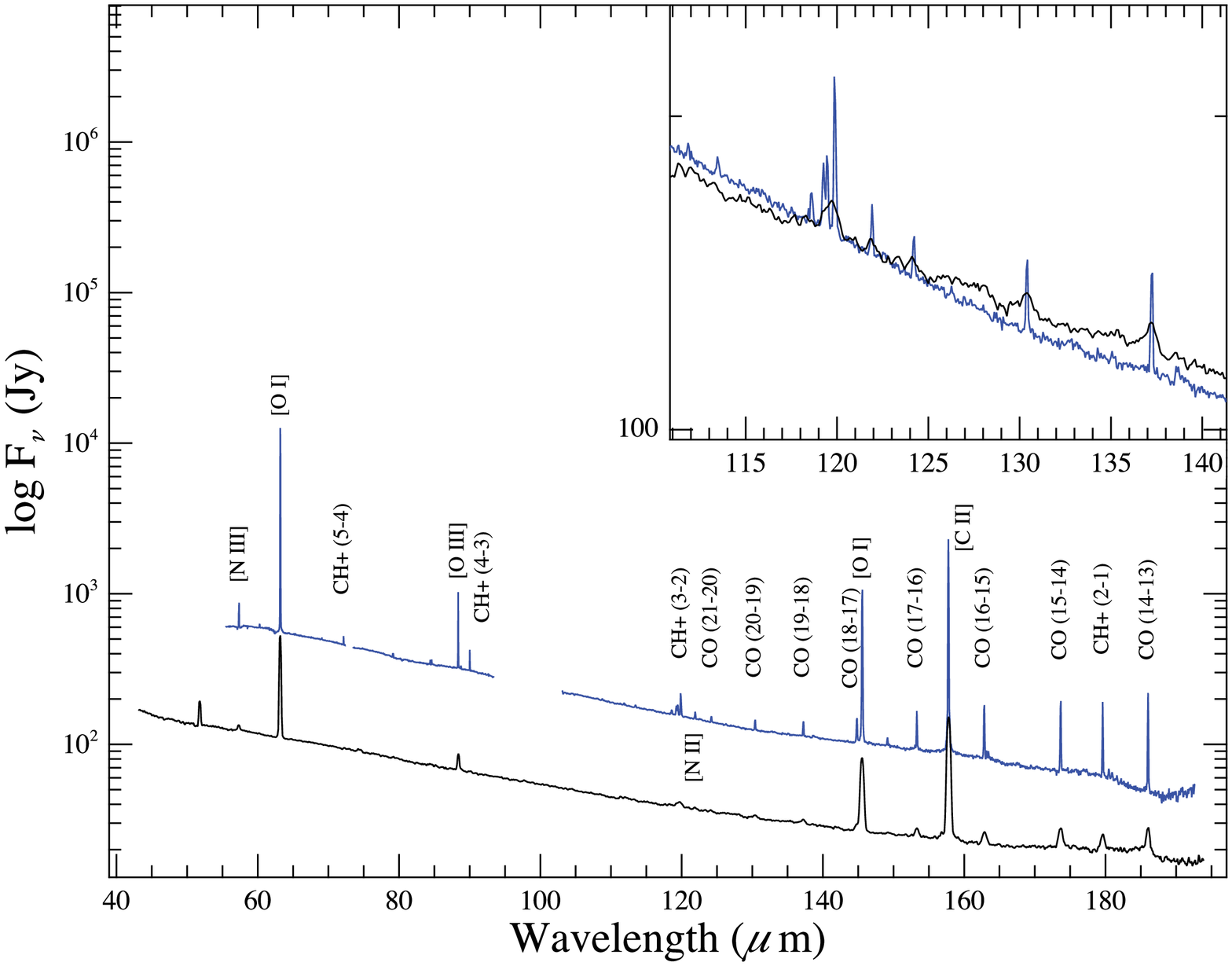}}
\end{minipage}

\caption[]{ 
Top panel:
SPIRE FTS spectrum of NGC 7027 (Wesson et al. 2010b), 
with $^{12}$CO, $^{13}$CO and CH$^+$ lines indicated.
Bottom Panel:
PACS spectrum of NGC 7027 (Exter et al. 2011; the blue upper curve) with the strongest lines indicated, 
and the ISO LWS spectrum (black, lower curve) of Liu et al. (1996) scaled by a factor of 0.25 in flux.
The inset shows a detail of the spectrum (with no scaling in flux) illustrating the improved resolution 
of the PACS spectrum w.r.t. the LWS spectrum.

} 
\label{Fig-colines} 
\end{figure}

That water will be a main theme is also illustrated in Royer et al. (2010) 
who present PACS and SPIRE spectroscopy of the RSG VY CMa. Nine hundred 
lines are identified, of which half are of water. 
Finally, Wesson et al. (2010b) present SPIRE FTS spectra of three
archetypal carbon-rich P-AGB objects, AFGL 618, AFGL 2688 and NGC 7027,
with many emission lines detected. Results include the first detection of 
water in AFGL 2688 and the detection of the fundamental $J$= 1-0 line
of CH$^+$ in the spectrum of NGC 7027 (see Fig.~\ref{Fig-colines}).
Figure~\ref{Fig-colines} also shows the PACS spectrum and for comparison the ISO LWS spectrum
in Liu et al. (1996) scaled by a factor of 0.25 in flux. 
The inset more clearly illustrates the improved sensitivity and spectral resolution over a smaller wavelength region.
The width of the LWS resolution element was 0.29 or 0.6 $\mu$m (depending on wavelength), while that of PACS varies between 0.03 and 0.12 $\mu$m.
The LWS effective aperture was 80\arcsec\ diameter, while the PACS spectrum is based on the central 3 $\times$ 3 spatial pixels, corresponding to 
28 $\times$ 28\arcsec.
The continuum shape of the PACS spectrum may still affected by the
fact the ground-based RSRF (relative spectral response functions) was
used. The noise appears to vary across the spectrum because the red
and blue detectors have different characteristics.
A full analysis of the PACS spectrum, including a detailed comparison
to the LWS spectrum will be presented elsewhere (Exter et al. 2011).

\subsection{Dust Spectroscopy}

De Vries et al. (2011) will present the first result on dust
spectroscopy, the detection of the 69 $\mu$m Forsterite feature in the
P-AGB star HD 161796. Identifying dust features has proven to be
perhaps not as easy as originally believed. Reasons are that a more
accurate determination of the PACS and SPIRE RSRFs and improved data reduction techniques are still
ongoing, and also the wealth of molecular and fine-structure lines that need to be
removed before relatively broad dust features can be identified.

\subsection{AGB and post-AGB stellar fluxes and a comparison to literature data}

The imaging observations of the AGB and P-AGB stars in the MESS
program have been largely completed for SPIRE and 65 percent
completed for PACS, and in this section we present the fluxes of the
central objects.
The fluxes are calculated by aperture photometry. The size of the
aperture is calculated in an automated way in a similar way as the
determination of the {\it sourceradius} described in Appendix~B:
The rms in the final map is determined, and then the 15, 12, 8, 12, 10, 8 $\sigma$ 
contours (for PACS 70, 100, 160, SPIRE 250, 350, 500 $\mu$m, respectively) 
are approximated by a circle. This radius of this circle can be
multiplied by a user-supplied value. The default value is 1.4 so that
the aperture includes all emission above approximately $3\sigma$.
For objects with detached shells or extended emission due to wind-ISM
interaction the aperture was changed manually to include as best as possible the central
object only (see the remarks in Table~\ref{Tab-flux}).
The location of the sky annulus for subtraction of the background was
verified manually to be as much as possible free of extended emission.

We tuned our automatic procedure of choosing the apertures on PACS
data of the asteroid Vesta taken on OD 160 during PV and the empirical
beam profiles provided by the SPIRE team.
We find apertures of 28 (PACS 70~$\mu$m) and 36 (PACS 100~$\mu$m) arcsecond for which the
PACS Scan Map Release Note states that they enclose about 97\% of energy. 
For the three SPIRE bands we find radii of 64, 80 and 118\arcsec. The ratio of 
this size to the PSF size is similar to that for PACS, suggesting that 
these radii also include about 97\% of energy.

Table~\ref{Tab-flux} lists the fluxes of the central source for the AGB and P-AGB stars observed 
by October 17th. 
Objects where the fluxes of the central object may still include emission from extended emission are flagged.
Listed are the flux and the aperture used between parenthesis. 
The formal error on the flux is negligible and almost always below
0.01 Jy (PACS), and 0.02 Jy (SPIRE).  The absolute flux calibration
accuracy of the PACS and SPIRE photometry is estimated to be 10-15\%
(Poglitsch et al. 2010, Griffin et al. 2010).

\begin{figure}
\includegraphics[width=68mm, angle=-90]{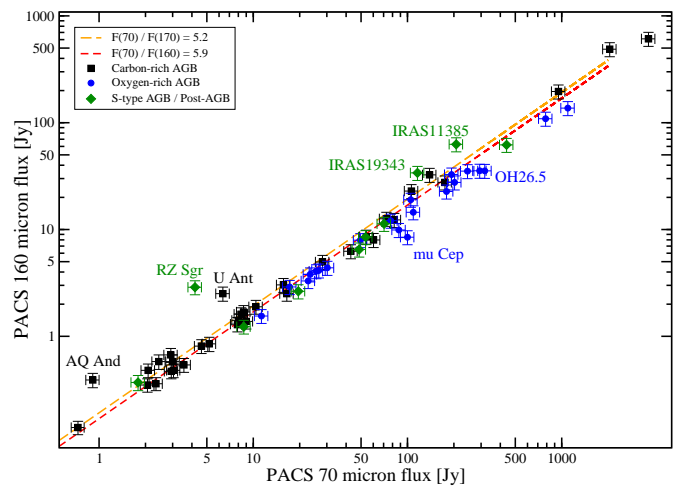}
\caption{PACS 70~$\mu$m versus PACS 170~$\mu$m flux densities for the observed targets in Table~2.}.
\label{Fig-pacs70-170}
\end{figure}

To make a first consistency check of the derived photometry we compare
the PACS 70 and 160~$\mu$m data with each other. The mean flux ratio
between 70 and 160~$\mu$m of the observed sources is 6.1, with
standard deviation of 1.5, which is a little higher, but within
the standard uncertainty, than the value of 5.2 expected for the ratio 
in fluxes of a modified black body 
($\nu^{\beta}B_{\nu}(T)$, with $\beta = 2$) on the Rayleigh-Jeans tail at the
observed wavelengths of 70 and 160~$\mu$m. However, the scatter is 
relatively large and for
13 out of the 68 sources the flux ratio deviates more than 30\%
(1$\sigma$) from the mean. The targets which have a significantly
lower flux ratio are all carbon-rich AGB or P-AGB, while the stars
with higher ratios are all oxygen-rich AGB (and 1 oxygen rich P-AGB). 
If we split the oxygen-rich and carbon rich stars in two groups, the
mean 70/160 flux ratios are 7.3 $\pm$ 1.4 and 5.5 $\pm$ 1.2, respectively.
Although both are consistent with the expected value, 
these ratios could indicate a larger value of $\beta$ for the
oxygen-rich stars than for the carbon-rich stars. 
These effects are illustrated in Fig.~\ref{Fig-pacs70-170}.
On the other hand a systematic difference in the dust temperature could alter 
these ratios as well, in particular if the spectral energy distribution deviates
from the Rayleigh-Jeans approximation at these wavelengths.

\begin{figure} 
\includegraphics[width=68mm,angle=-90]{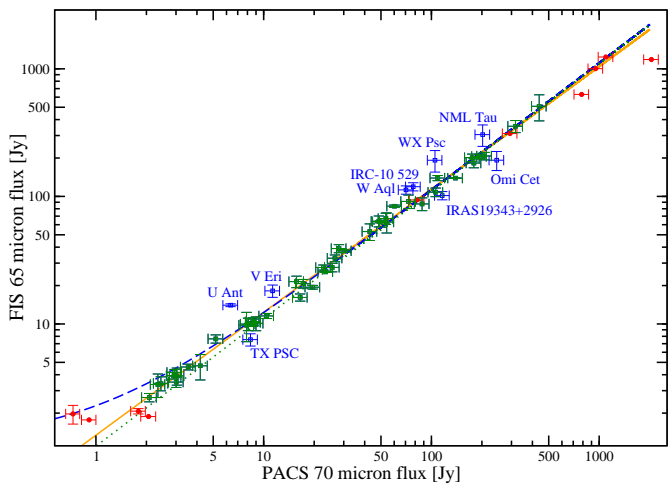}
\caption[]{%
{\it AKARI/FIS} 65~$\mu$m flux density plotted against PACS
  70~$\mu$m flux density.  The blue dashed line shows the best linear
  regression, after 3$\sigma$ outlier (blue open squares) rejection
  ($y=1.3+1.1\cdot x$), while the orange solid lines is the best fit
  using a power law relation ($y=1.3~x^{0.96}$).  The former fits were
  not constrained to go through the origin.  The dotted green line
  represents the nominal flux ratio of an A0 star for effective
  wavelengths of 62.8 and 68.8 for {\it FIS} and PACS,
  respectively. Objects indicated with red filled circles were omitted
  from the fit due to low quality flag in the {\it FIS} catalog.}  
\label{Fig-pacs-fis} 
\end{figure}

\begin{figure}
\includegraphics[width=68mm, angle=-90]{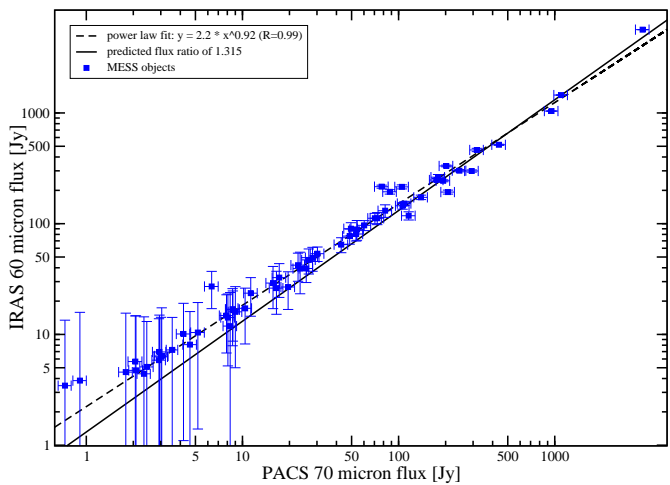}
\caption{PACS 70~$\mu$m versus IRAS 60~$\mu$m flux densities for
  the 62 observed targets in Table~2 with counterparts in the IRAS
  point source catalog. The solid line shows the predicted flux ratio
  of 1.3 (see text) and the dashed line the powerlaw fit to all
  objects brighter than 10~Jy.}
\label{Fig-pacs-iras}
\end{figure}

A second consistency and data quality check of the measured PACS central
source flux densities is performed by cross-matching targets and
comparing fluxes in this work with the {\it AKARI/FIS} bright source catalog
(Yamamura et al. 2010). We find 59 matches with the {\it FIS} bright source
catalog, of which 49 have good quality 65~$\mu$m data (the low quality
data correspond to {\it FIS} sources brighter than 500~Jy or fainter than
$\sim$2~Jy). The correlation coefficient for these two data sets  (omitting 
the bad quality {\it FIS} flux densities) is very
good with $R$= 0.98 (see Fig.~\ref{Fig-pacs-fis}). The regression improves 
(in particular at lower flux levels) from choosing a power law instead of a 
linear function. 
The slope of the linear function is close to expected  
flux ratio of 1.1 (again for a modified black body with $\beta = 2$).
NML Tau and $o$ Cet deviate significantly from this 
relation but these sources are relatively bright ($>$200 Jy) and may
therefore suffer from inaccurate {\it FIS} fluxes. Other sources deviating
from the linear relationship (with {\it FIS} fluxes higher than PACS fluxes)
are NML Tau, WX Psc, and U Ant (see below).
A similar comparison of PACS 70~$\mu$m flux densities with IRAS 60~$\mu$m 
values is shown in Fig.~\ref{Fig-pacs-iras}.
We note that in general flux densities appear consistent with previous results
thus warranting a direct comparison for SED modelling. However in a few cases
extended or detached emission near the AGB stars contaminated previous flux measurements 
of the central source (see also below for examples), 
thus introducing errors in the results obtained with SED models.
Additional errors could arise from flux variability on time-scales of a few years up to two decades.
Also, errors in the automated point source extraction routines for both IRAS and AKARI could
have introduced unknown errors for particular objects.

A comparison between the PACS 160~$\mu$m and {\it FIS} 140 or 160~$\mu$m flux
densities proved less conclusive due to the relative large uncertainties
on the {\it FIS} fluxes (these are not reliable for reported fluxes below $\sim$3 Jy, 
nor above 100 Jy). Provisionally, the PACS 160~$\mu$m fluxes are
consistent with previous measurements though in several cases we noticed
that the measured values are higher than expected from the overall SED shape. 
Again, the presence of strong emission lines could affect the broad band flux measurements.

We discuss a few particular cases from the comparisons made above.
For the S-type star RZ Sgr the 70~$\mu$m flux is only 1.5 times lower than
at 160~$\mu$m. It is likely that the latter value is still contaminated by extended emission in the aperture. 
The 70~$\mu$m flux (4.2 $\pm$ 0.4~Jy) is consistent with the 65~$\mu$m {\it FIS} flux  (4.7 $\pm$ 1.1~Jy). 
Also, the {\it FIS} fluxes at 140 and 160~$\mu$m are somewhat higher, though of low quality, 
than the PACS 160~$\mu$m flux (i.e. $\sim$4 - 5 Jy versus  $\sim$2.9~Jy).
Likely, the {\it FIS} aperture is affected by the extended emission seen in the PACS red image.
TT Cyg also has an average 70/160~$\mu$m flux ratio.
The central source flux at 70~$\mu$m reported for PACS is a third of the {\it FIS} 65~$\mu$m flux. 
Here the PACS 70~$\mu$m map shows a bright detached shell. Therefore the {\it FIS} flux could include 
both the central source and detached shell. Additionally, the {\it FIS} 65~$\mu$m flux
has a low-quality flag.
In this case the {\it FIS} fluxes at 140 and 160~$\mu$m (i.e. 1.5 and 0.3 Jy) 
agree better with the 160~$\mu$m flux of 0.14 Jy. The detached shell is fainter at 
160~$\mu$m and thus the difference between PACS and {\it FIS} will be less.
The PACS 70 and 160~$\mu$m point source fluxes 
for U Ant (6.3 and 2.5 Jy, respectively) are significantly lower than the corresponding 
{\it FIS} fluxes of 14.0, 14.0, 5.8 and 4.8 Jy at 65, 90, 140, and 160~$\mu$m.
In this case we surmise that the bright extended structure visible in
the PACS images (which is not included in the point source flux) is
included in the point source flux reported in both the {\it FIS} and IRAS catalogs.
For AQ And, which has a bright detached shell in both PACS images, the
central source is a factor two to three fainter than the corresponding
fluxes reported in the {\it FIS} catalog. However, for this faint source the
{\it FIS} values have a low quality flag. Nevertheless, it seems likely that
the latter include (at least part of) the shell emission as resolved by PACS.
These examples are primarily given to illustrate the complexities involved when comparing
results from different surveys/instruments and in particular when those results are
used to construct and model spectral energy distributions.
The latter will benefit also from additional information provided by previous missions 
such as ISO and Spitzer, as well as PACS spectroscopy observations of some MESS targets (Table 1).

\begin{table*}
\setlength{\tabcolsep}{1.3mm}

\caption{Stellar fluxes (in Jy) in the apertures listed between parenthesis (in arcseconds) 
of the AGB and post-AGB stars observed per October 17th. 
Add 1342100000 to the AOR number quoted to obtain the Observation ID. 
Date refers to the observing date (format yy-mm-dd), when two are listed the first refers 
to the PACS observations, the second to the date of the SPIRE observations.}

\begin{tabular}{lrrrrrlll}
\hline
        Object &  PACS blue     & red        & SPIRE PSW     & PMW        & PLW  & AOR & Date & Remarks \\
               &     70 $\mu$m   &  160 $\mu$m &  250 $\mu$m    &  350 $\mu$m &  500 $\mu$m &   \\
\hline
        AFGL 190 &  42.9 (33) &  6.26 (31) & &&& 89183, 89184  & 10-01-12  \\
       AFGL 2155 &  54.1 (35) &  8.57 (34) &  &&& 92777, 92778  & 10-03-26  \\
       AFGL 2688 &  2047 (110) &  488 (118) &  119 (106) &  48.9 (140) &  18.2 (92) & 95837, 95838, 88168  & 10-05-05, 09-12-16  \\
        AFGL 278 &  22.7 (29) &  3.30 (28) &  &&& 89840, 89841  & 10-01-28  \\
       AFGL 3068 &  175 (40) &  27.6 (34) &  6.93 (56) &  3.32 (136) &  1.26 (92) & 88178, 88378, 88379  & 09-12-21, 09-12-16  \\
       AFGL 3116 &  73.0 (46) &  12.6 (50) &  &&& 88490, 88491  & 09-12-23  \\
        AFGL 618 &  953 (104) &  196 (86) &  46.1 (101) &  19.0 (87) &  8.91 (92) & 93132, 93133, 91183  & 10-03-31, 10-02-25  \\
    $\alpha$ ORI &  294 (130) &  35.5 (82) &  6.18 (57) &  2.84 (68) &  1.56 (92) & 104435, 104436, 92099  & 10-09-13, 10-03-11 & c  \\
          AQ AND &  0.914 (14) &  0.391 (28) & &&& 88488, 88489  & 09-12-23 & a \\ 
          AQ SGR &  2.05 (14) &  0.354 (28) & &&& 95702, 95703  & 10-04-30  \\
           CIT 6 &  & & 8.81 (72) &  3.77 (85) &  1.77 (92) & 106689  & 10-10-17  \\

          CW LEO &  3653 (244) &  610 (253) &  138 (211) &  57.7 (325) &  23.2 (276) & 86298, 86299, 86293  & 09-10-25 \\

          EP AQR &  26.6 (31) &  4.14 (31) & &&& 95460, 95461  & 10-04-23 & b \\
       HD 161796 &  109 (42) &  14.5 (44) &  2.39 (45) &  0.731 (63) &  0.254 (92) & 88155, 91127, 91128  & 10-02-24, 09-12-15  \\
       HD 179821 &  439 (63) &  62.0 (55) & &&& 93490, 93491  & 10-04-03  \\
       HD 235858 &  60.2 (42) &  7.98 (28) & &&& 96757, 96758  & 10-05-20 & c \\
        HD 44179 &  139 (49) &  32.5 (55) & 9.51 (50) &  4.46 (67) &  2.46 (92) & 85549, 85550, 83682  & 09-10-09, 09-09-12  \\
        HD 56126 &  & & 1.21 (45) &  0.190 (92) &  0.388 (63) & 93007  & 10-03-28  \\
 IRAS 11385-5517 &  207 (55) &  62.8 (62) & &&& 102339, 102340  & 10-08-09  \\
 IRAS 15194-5115 &  106 (48) &  22.9 (72) & &&& 90247, 90248  & 10-02-05  \\
 IRAS 16594-4656 &  82.0 (39) &  12.3 (28) & &&& 93052, 93053  & 10-03-30  \\
 IRAS 19343+2926 &  116 (44) &  33.9 (46) & &&& 91951, 91952  & 10-03-10  \\
     IRC -10 529 &  77.9 (40) &  12.1 (44) &  &&& 96034, 96035  & 10-05-09  \\
      $\chi$ CYG &  53.5 (48) &  8.49 (38) & &&& 88320, 88321  & 09-12-20  \\
       $\mu$ CEP &  100 (86) &  8.47 (39) &  &&& 91945, 91946  & 10-03-10  & b\\
         NML CYG &  786 (82) &  109 (35) & &&& 95485, 95486  & 10-04-24 & c \\
         NML TAU &  202 (69) &  27.6 (67) &  5.99 (63) &  2.71 (119) &  0.876 (92) & 90343, 90344, 91180  & 10-02-10, 10-02-25 & c \\
    OH 127.8+0.0 &  87.9 (41) &  9.90 (31) & &&& 89181, 89182  & 10-01-12  \\
     OH 26.5+0.6 &  317 (44) &  35.4 (28) & &&& 91817, 91818  & 10-03-09 & c \\
         $o$ CET &  245 (99) &  35.2 (90) & 6.87 (63) &  2.71 (73) &  1.27 (92) & 90335, 90336, 89423  & 10-02-09, 10-01-17 & c \\
         $o$ ORI &  1.79 (14) &  0.368 (28) & &&& 91108, 91109  & 10-02-23  \\
       $\pi$ GRU &  48.6 (56) &  6.50 (47) & 1.54 (45) &  0.687 (63) &  0.313 (92) & 96799, 96800,93791  & 10-05-21, 10-04-05 & c \\
           R AQR &  & & 1.03 (45) &  0.423 (63) &  0.226 (92) & 88166  & 09-12-16  \\
           R CAS &  & & 2.90 (45) &  1.51 (63) &  0.917 (92) & 88578  & 09-12-25 & c \\
           R DOR &  193 (79) &  32.7 (57) & 9.63 (62) &  4.73 (154) &  2.19 (92) & 97685, 97686, 88164  & 10-06-03, 09-12-16  \\
           R FOR &  9.00 (21) &  1.38 (28) &  &&& 88502, 88503  & 09-12-24  \\
           R HYA &  49.5 (37) &  7.95 (32) &  2.88 (47) &  1.50 (85) &  0.841 (92) & 102231, 102232, 88662  & 10-08-06, 09-12-27  \\
           R LEP &  16.5 (28) &  2.51 (28) &  &&& 90304, 90305  & 10-02-07  \\
           R SCL &  & & 1.63 (45) &  0.692 (63) &  0.303 (92) & 88657  & 09-12-26  \\
          RT CAP &  2.34 (16) &  0.363 (28) & &&&  96036, 96037  & 10-05-09  \\
          RT VIR &  23.4 (39) &  3.80 (34) &  &&& 101181, 101182  & 10-07-24  \\
          RX BOO &  & & 1.44 (25) &  0.798 (63) &  0.402 (92) & 88158,  & 09-12-16  \\
          RY DRA &  4.61 (19) &  0.810 (28) & &&&  88338, 88339  & 09-12-21  \\
          RZ SGR &  4.19 (22) &  2.88 (31) &  &&& 93532, 93533  & 10-04-03 & b \\

\hline
\end{tabular}
\end{table*}

\setcounter{table}{1} 
\begin{table*}
\setlength{\tabcolsep}{1.9mm}

\caption{Continued }

\begin{tabular}{lrrrrrlll}
\hline
        Object &  PACS blue     & red       & SPIRE PSW     & PMW     & PLW  & AOR & Date & Remarks \\
               &     70 $\mu$m   &  160 $\mu$m &  250 $\mu$m    &  350 $\mu$m &  500 $\mu$m &   \\
\hline
           S CAS &  19.6 (26) &  2.64 (28) & &&&  105040, 105041  & 10-09-22  \\
           S CEP &  15.7 (27) &  3.03 (28) &  1.09 (45) &  0.759 (63) &  0.410 (92) & 88342, 88343, 103604 & 09-12-21, 10-08-24 & a \\
          ST CAM &  3.04 (16) &  0.477 (28) &  &&& 106044, 106045  & 10-10-07  \\
          ST HER &  8.68 (26) &  1.23 (28) &  &&& 88324, 88325  & 09-12-20  \\
    $\theta$ APS &  30.2 (37) &  4.38 (33) &  1.05 (45) &  0.476 (63) &  0.174 (92) & 93048, 93049, 103637  & 10-03-30, 10-08-24 & a \\
           T LYR &  2.92 (15) &  0.673 (28) &  &&& 95694, 95695  & 10-04-30  \\
           T MIC &  17.2 (26) &  2.90 (28) &  &&& 93036, 93037  & 10-03-30 & c \\
          TT CYG &  0.725 (14) &  0.139 (14) & &&& 93500, 93501 & 10-04-03 & a \\
          TW HOR &  3.55 (18) &  0.537 (28) &  &&& 88373, 88374  & 09-12-21  \\
          TX PSC &  8.34 (43) &  1.61 (35) &  &&& 88344, 88345 & 09-12-21 & b \\
           U ANT &  6.35 (14) &  2.51 (28) & 0.491 (25) &  0.210 (25) &  0.074 (25) & 88474, 88475, 88161  & 09-12-23, 09-12-16  & d \\
           U CAM &  8.65 (21) &  1.69 (28) & 0.891 (45) &  0.515 (63) &  0.239 (92) & 90307, 90308, 103608  & 10-02-07, 10-08-24  \\
          UU AUR &  8.52 (26) &  1.44 (28) &  &&& 106332, 106333  & 10-10-12  \\
          UX DRA &  2.08 (17) &  0.481 (28) & 0.156 (45) &  0.110 (63) &  0.099 (92) & 88340, 88341, 103603  & 09-12-21, 10-08-24 & a \\
           V CRB &  3.01 (17) &  0.580 (28) & &&&  88326, 88327  & 09-12-20  \\
           V CYG &  28.1 (37) &  4.97 (28) & &&&  88462, 88463  & 09-12-22 & b \\
           V ERI &  11.3 (27) &  1.55 (28) &  &&& 89943, 89944  & 10-01-31  \\
           V HYA &  & & 2.74 (50) &  1.31 (63) &  0.552 (92) & 88160  & 09-12-16  \\
           V PAV &  5.18 (19) &  0.845 (28) &  &&& 104278, 104279  & 10-09-10  \\
          VX SGR &  179 (62) &  22.7 (36) &  &&& 91799, 91800  & 10-03-09  \\
          VY CMA &  1096 (102) &  137 (28) &  &&& 94070, 94071  & 10-04-07  \\
           W AQL &  70.5 (58) &  11.3 (59) &  &&& 94084, 94085  & 10-04-08 & b \\
           W HYA &  & & 7.50 (49) &  3.39 (67) &  1.66 (92) & 89519  & 10-01-19 & c \\
           W ORI &  8.07 (22) &  1.41 (28) &  &&& 90965, 90966  & 10-02-22  \\
           W PIC &  2.44 (14) &  0.575 (28) &  &&& 90974, 90975  & 10-02-22  \\
          WX PSC &  105 (50) &  19.0 (68) &  &&& 88486, 88487  & 09-12-23  \\
           X HER &  25.6 (47) &  4.07 (40) &  &&& 88322, 88323  & 09-12-20 & b \\
           X TRA &  7.90 (25) &  1.30 (28) &  &&& 93050, 93051  & 10-03-30  \\
           Y CVN &  10.4 (27) &  1.89 (28) & 0.630 (45) &  0.468 (63) &  0.235 (92) & 88330, 88331, 88157 & 09-12-20, 09-12-15 & b \\
           Y PAV &  2.94 (16) &  0.468 (28) &  &&& 88157  & 09-12-15  \\

\hline
\end{tabular}

Remarks:
Sources are listed alphabetically.
Internal errors in the fluxes are small and of order 0.01 Jy (PACS) and 0.02 Jy (SPIRE).
The errors in the fluxes are dominated by the flux calibration accuracy, currently estimated to be 
10\% for the PACS 70 $\mu$m band and better than 15\% at 160 $\mu$m, and 15\% for all SPIRE bands.

(a) Aperture includes only the central star; 
(b) Aperture includes (part of) the extended emission; 
(c) Some contamination by extended emission in the aperture or sky annulus; 
(d) Only PACS blue and red and SPIRE PSW are essentially free from emission of the detached shell.

\label{Tab-flux}
\end{table*}

\section{Conclusions and outlook} 

The scope, aims and status of the \it Herschel \rm Guaranteed Time Key Program MESS (Mass-loss of Evolved StarS) are presented.
The current concepts on the data processing are presented, and
aperture photometry of all 70 AGB and post-AGB stars observed as per
October 17th 2010 are 
presented\footnote{The progress of the MESS project can be followed via www.univie.ac.at/space/MESS}.

Some of our SDP data is already public (see the remarks in
Table~\ref{list-all}) and the data taken in routine phase have a
proprietary period of one year, implying that data will become
successively public from about December 2010 onwards.

Currently, the PACS maps are produced via the standard {\tt PhotProject} task, that 
use data that are filtered to remove 1/f-noise as input. 
We achieve good results for compact objects, but unmasked large-scale 
structures in the background are affected by the filtering. In order to 
improve this, we are currently investigating other filtering and mapping 
methods, such as MADmap (Cantalupo et al. 2010). That way we also seek 
to improve the spatial resolution of the final maps.
A second point under investigation is to correct the effects caused by 
the instrument PSF. With its tri-lobe pattern and other wide-stretched 
features it is currently not possible to make any definite statements 
about structure in the circumstellar emission close to the central 
object, although many sources are extended. Thus we are investigating 
different deconvolution strategies and PSF-related matters.
The efforts the MESS consortium are currently undertaking to improve the 
PACS data processing are described in Ottensamer et al. (2011).

On the science side, the publication of the very first {\it Nature} paper
based on {\it Herschel} results by Decin et al. (2010b) is a highlight.
It demonstrates the power of the PACS and SPIRE spectrometers. 
With more than 50 PACS and almost 30 SPIRE targets to be observed spectroscopically 
this will result in an extremely rich database that, with proper modelling, will allow 
detailed studies on molecular abundances, the velocity structure in the
acceleration zone close to the star, and the mass-loss rate.

On the imaging side the fact that bow shock cases are ubiquitous is extremely interesting.
Although this in fact makes it more difficult to derive the mass-loss rate history of the AGB star, 
it offers an unique opportunity to use these cases as probes of the ISM.

\acknowledgements{  
%
JB, LD, KE, AJ, MG, PR, GvdS, SVE, PvH, C.V-N.and BvdB acknowledge support from the 
Belgian Federal Science Policy Office via de PRODEX Programme of ESA.
SVE, Y.N. and D.H. are supported by the F.R.S-FNRS. 
FK and RO acknowledge funding by the Austrian Science Fund FWF under project number I163-N16.
PACS has been developed by a consortium of institutes led by MPE
(Germany) and including UVIE (Austria); KUL, CSL, IMEC (Belgium); CEA,
OAMP (France); MPIA (Germany); IFSI, OAP/AOT, OAA/CAISMI, LENS, SISSA
(Italy); IAC (Spain). This development has been supported by the funding
agencies BMVIT (Austria), ESA-PRODEX (Belgium), CEA/CNES (France),
DLR (Germany), ASI (Italy), and CICT/MCT (Spain).
SPIRE has been developed by a consortium of institutes led by
Cardiff Univ. (UK) and including Univ. Lethbridge (Canada);
NAOC (China); CEA, LAM (France); IFSI, Univ. Padua (Italy);
IAC (Spain); Stockholm Observatory (Sweden); Imperial College London,
RAL, UCL-MSSL, UKATC, Univ. Sussex (UK); and Caltech, JPL, NHSC,
Univ. Colorado (USA). This development has been supported by
national funding agencies: CSA (Canada); NAOC (China); CEA,
CNES, CNRS (France); ASI (Italy); MCINN (Spain); SNSB (Sweden); STFC (UK); and NASA (USA).
HSpot is a joint development by the Herschel Science Ground Segment
Consortium, consisting of ESA, the NASA Herschel Science Center, and
the HIFI, PACS and SPIRE consortia.
This research is based on observations with {\it AKARI}, a JAXA project with
the participation of ESA.
}

{}

\newpage

\begin{appendix}
\section{Lessons from the SDP}

\begin{figure} 

\begin{minipage}{0.22\textwidth}
\resizebox{\hsize}{!}{\includegraphics{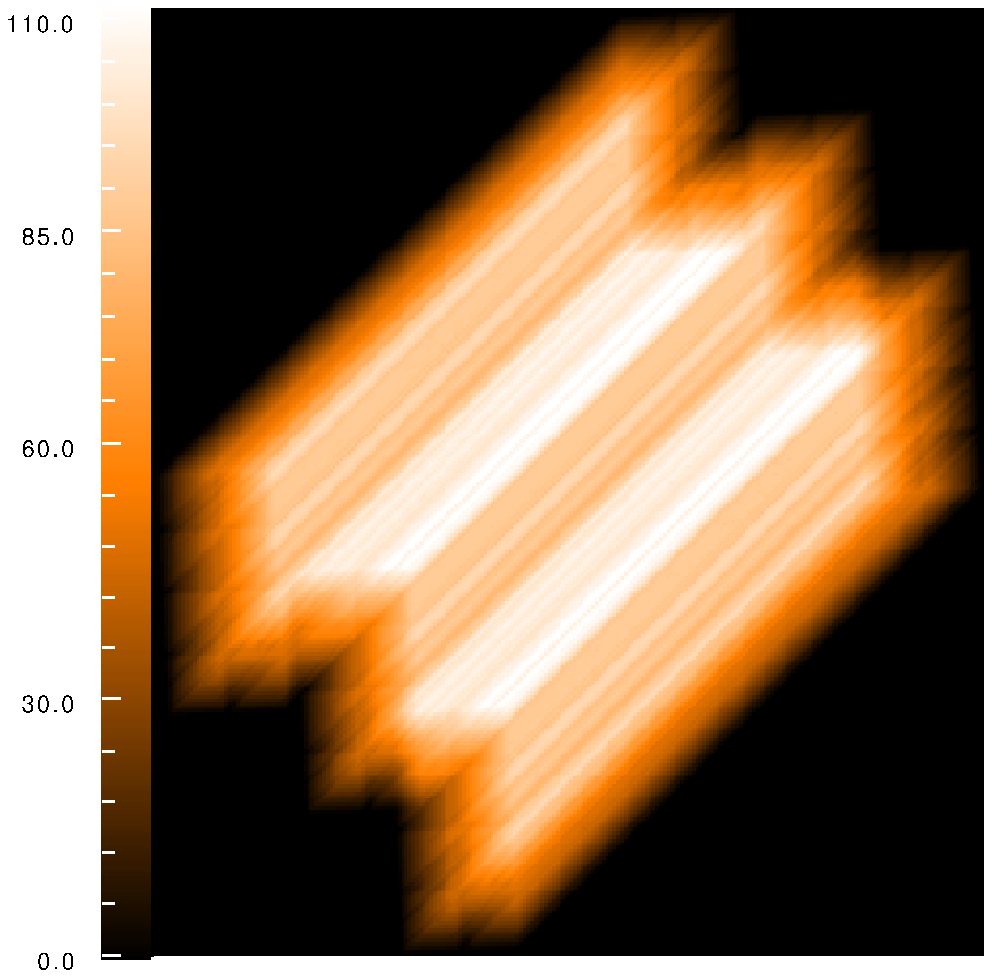}}
\end{minipage}
\begin{minipage}{0.22\textwidth}
\resizebox{\hsize}{!}{\includegraphics{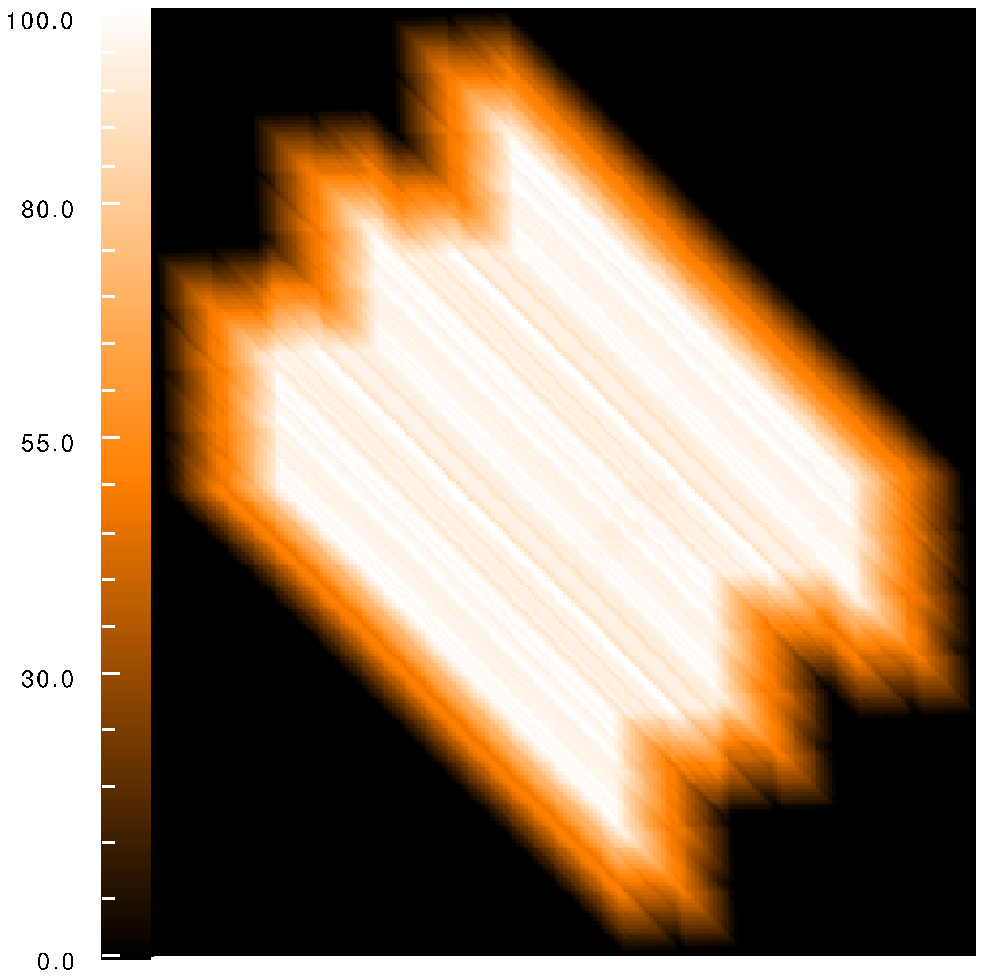}}
\end{minipage}

\begin{minipage}{0.22\textwidth}
\resizebox{\hsize}{!}{\includegraphics{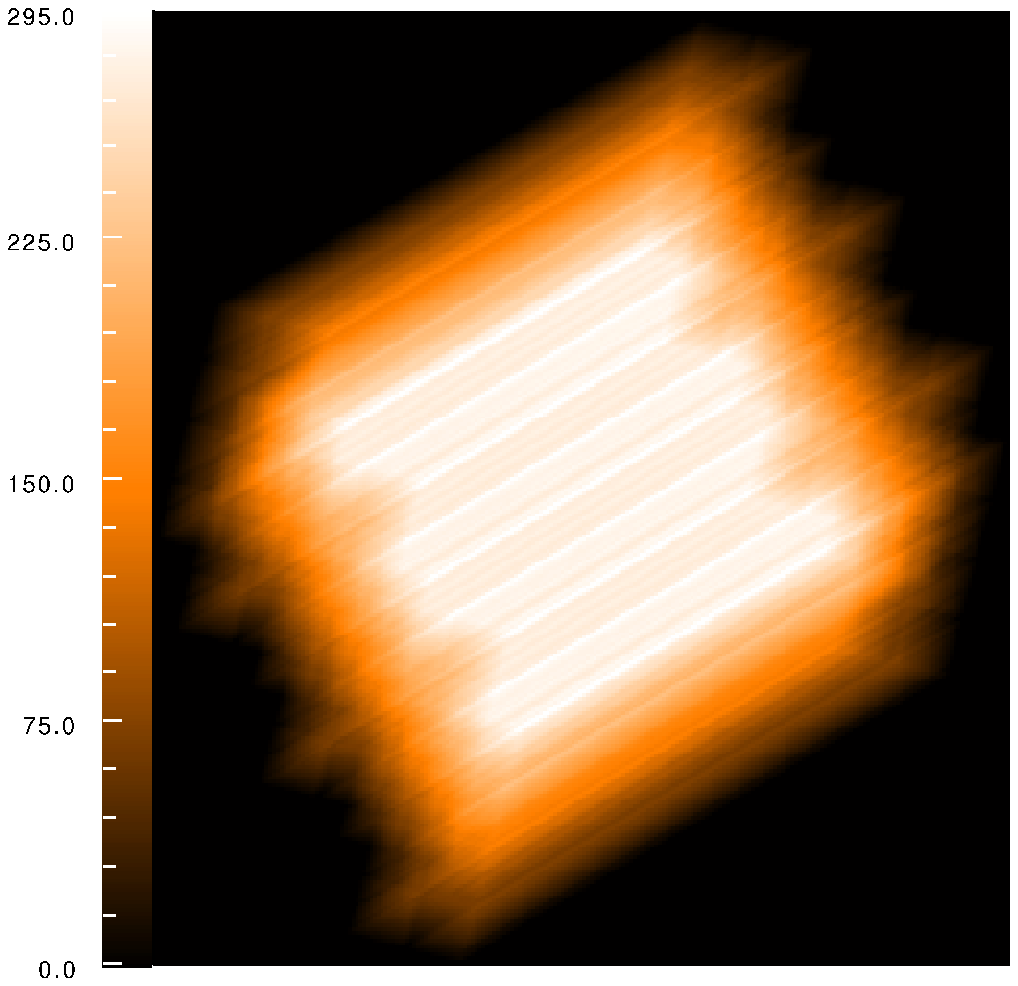}}
\end{minipage}
\begin{minipage}{0.22\textwidth}
\resizebox{\hsize}{!}{\includegraphics{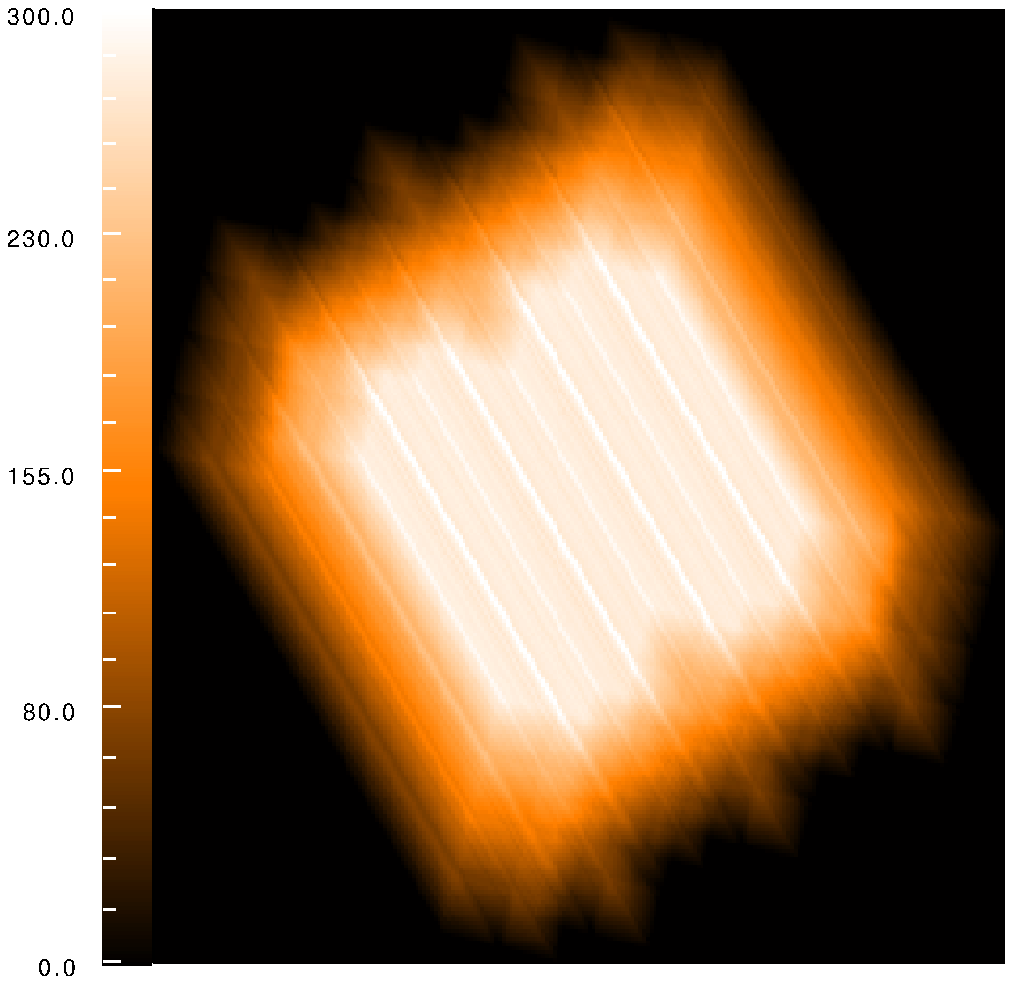}}
\end{minipage}

\caption[]{ 
Top:    Coverage map of the scan and cross-scan (left and right) of TT Cyg obtained in SDP on OD 148.
Bottom: Coverage map of the scan and cross-scan of AQ And obtained in Routine Phase on OD 224.
Until OD 221 the coverage of the scan map obtained at an angle of 45\degr\ was not uniform.
} 
\label{Fig-cov} 
\end{figure}

A few sources were proposed to be observed during Science Demonstration Phase (SDP).
This was intended as a demonstration of the implementation of the concatenation of a scan
and a cross-scan, at the recommended array-to-map angles of 45 and 135\degr.
It was discovered that the coverage of maps obtained at 45\degr\ was not uniform. 
In the example shown in top left panel of Fig.~\ref{Fig-cov} the coverage in the center 
of the map is only about 80\% of the maximum coverage.
This was reported to the instrument team. The plate scale of the PACS
camera was then recalibrated, which led to a 3\% larger pixel size on
the sky. An extra rotation of 2.5 degrees of the instrument coordinate
system w.r.t. the satellite reference also had to be introduced.
The changes have been implemented on OD 221, and
since then the coverage is now much more uniform (see lower panel of Fig.~\ref{Fig-cov}).
No changes are visible to the user, and so in HSPOT angles of 45 and 135 degrees still need to be entered.

Originally we opted to use ``homogeneous coverage'' and ``square map''
as observing mode parameters for the scan maps.  In that case the
HSPOT internal logic takes the scan-leg length from the user input to
compute the cross-scan step and the number of scan-legs needed.
It turned out however that a small change in scan-leg length can then
make a large difference in the computed observing time. Therefore, we
have computed the discrete set of optimum solutions for the scan-leg
length and cross-scan distance corresponding to the square maps
obtained from perpendicular scans performed under array-to-map angles
of 45 and 135 degrees. The standard cross-scan distance for uniform
coverage is 155 arcsec. For repetitions of 2 or 4, this can be divided
accordingly, and intermediate solutions are found (see Table~\ref{Tab-magic}), 
which are more efficient than repeating the whole scan.
For example, if one would like to cover an area of 60\arcmin$\times$60\arcmin\ homogeneously, 
one can opt for a cross-scan step of 155\arcsec, a scan-leg length of 64.4\arcmin, and 25 scan legs.

\begin{table}

\caption{Scan map observing mode parameters}

\begin{tabular}{ccc}
\hline
number of scan legs & Uniformly Covered	& scan-leg  length \\
                    & ($\arcmin$)      &  ($\arcmin$)     \\
\hline
\multicolumn{3}{c}{cross-scan step= 155 \arcsec}  \\
1  &      1.24  &      4.95  \\
2  &      3.71  &      7.42   \\
3  &      6.19  &      9.90   \\
4  &      8.66  &     12.37   \\
5  &     11.14  &     14.85   \\
6  &     13.61  &     17.32   \\
7  &     16.09  &     19.80   \\
8  &     18.56  &     22.27   \\
9  &     21.04  &     24.75   \\
10  &     23.51  &     27.22   \\
11  &     25.99  &     29.70   \\
12  &     28.46  &     32.17   \\
13  &     30.94  &     34.65   \\
14  &     33.41  &     37.12   \\
15  &     35.89  &     39.60   \\
16  &     38.36  &     42.07   \\
17  &     40.84  &     44.55   \\
18  &     43.31  &     47.02   \\
19  &     45.79  &     49.50   \\
20  &     48.26  &     51.97   \\
21  &     50.73  &     54.45   \\
22  &     53.21  &     56.92   \\
23  &     55.68  &     59.40   \\
24  &     58.16  &     61.87   \\
25  &     60.63  &     64.35   \\
\multicolumn{3}{c}{cross-scan step= 77.5 \arcsec}  \\
 3  &      1.24  &      4.95  \\
 4  &      2.47  &      6.19  \\
 5  &      3.71  &      7.42  \\
 6  &      4.95  &      8.66  \\
 7  &      6.19  &      9.90  \\
 8  &      7.42  &     11.14  \\
 9  &      8.66  &     12.37  \\
10  &      9.90  &     13.61  \\
11  &     11.14  &     14.85  \\
12  &     12.37  &     16.09  \\
13  &     13.61  &     17.32  \\
14  &     14.85  &     18.56  \\
15  &     16.09  &     19.80  \\
16  &     17.32  &     21.04  \\
17  &     18.56  &     22.27  \\
18  &     19.80  &     23.51  \\
19  &     21.04  &     24.75  \\
20  &     22.27  &     25.99  \\
21  &     23.51  &     27.22  \\
22  &     24.75  &     28.46  \\
23  &     25.99  &     29.70  \\
24  &     27.22  &     30.94  \\
25  &     28.46  &     32.17  \\
26  &     29.70  &     33.41  \\
27  &     30.94  &     34.65  \\
28  &     32.17  &     35.89   \\
29  &     33.41  &     37.12  \\
30  &     34.65  &     38.36  \\
31  &     35.89  &     39.60  \\
32  &     37.12  &     40.84  \\
33  &     38.36  &     42.07  \\
34  &     39.60  &     43.31  \\
36  &     42.07  &     45.79  \\
38  &     44.55  &     48.26  \\
40  &     47.02  &     50.73  \\
42  &     49.50  &     53.21  \\
44  &     51.97  &     55.68  \\
46  &     54.45  &     58.16  \\
48  &     56.92  &     60.63  \\
50  &     59.40  &     63.11  \\

\hline
\end{tabular}

\label{Tab-magic}
\end{table}

\section{PACS image reduction}
In this Appendix the strategy that was used to reduce the PACS images is outlined in some detail.
It is mainly intended for persons who are interested in, or want to learn more about, PACS 
image reduction using the ``high pass filter'' technique. 

As mentioned in Sect.~\ref{pacs-ima} the data reduction starts from
``Level 0'' (raw data) as retrieved from the {\it Herschel} Science Archive.
Compared to the ``Level 2'' product as produced by execution of the
standard pipeline steps, as described by Poglitsch et al. (2010),
special care need to be given to the deglitching step(s), and the
``high pass filtering''. Therefore, the making of the map has become a
two-step process. In addition, as the standard pipeline script
operates on a single AOR, while our observations are always the
concatenation of 2 AORs (a scan and an orthogonal cross-scan) an additional step
is also needed.  All steps below are valid and tested under 
HIPE v4.4.0\footnote{Data presented in this paper were analysed using
  HIPE, a joint development by the Herschel Science Ground Segment
  Consortium, consisting of ESA, the NASA Herschel Science Center, and
  the HIFI, PACS and SPIRE consortia.  (See
  http://herschel.esac.esa.int/DpHipeContributors.shtml) } (Ott et al. 2010).

The generation of the level 1 product follow the description in Poglitsch et al.
In particular, the deglitching is done using the {\tt photMMTDeglitching} task
\footnote{
using 
photMMTDeglitching(frames, scales=3, nsigma=9, incr\_fact=2, mmt\_mode='multiply', onlyMask= True)
}
and frames detected as glitches are only masked without trying to interpolate the signal.
When the central source is bright however, then this task also incorrectly masks a significant number of 
frames of the central source.
Therefore a second step is required as detailed below.

In the case of a scan map AOR the level 2 product is a map. 
The processing steps from the level 1 product to this map are the {\tt highpassFilter} task and the {\tt photProject} task.

The purpose of the high pass filter is to remove detector drift and 1/f noise.  
At the moment the task is using a median filter, which subtracts a
running median from each readout. The filter box size can be set by the user. 
For a certain high pass filter width (HPFW), the median is run over
($2\cdot$HPFW+1) consecutive readouts, and only uses readouts that are not masked.
The filter works in the time domain. However this corresponds to
filtering on a certain spatial scale as the scan speed (20\arcsec/s in our case)
and the sampling of the photometer signal (10 Hz) are known.
In the making of the first map a HPFW of half of the scan-leg of the map is used.

The map making task simply reprojects the R.A. and Declination of every frame onto the sky using a certain pixel size.
For the generation of this first map the native pixel sizes are used (3.2\arcsec\ in the blue, and 6.4\arcsec\ in the red).

The map generated in this way (and essentially equivalent to the level 2 product 
retrieved from the HSA) is then analysed in order to make the second and final map for an AOR. 
In particular, the location of all significant emission in the map
will be determined, especially the location and extent of the central
source will be characterised. With this information the second level
deglitching can be performed, the significant emission in the map will
be masked, and the optimised value of the HPFW will be determined.

The importance of properly masking the significant emission in the map
and the choice of the HPFW (a related issue) will be illustrated in detail first. 
For this we use the observations of AQ And at 70$\mu$m.  This star has
a very nice detached shell and the PACS observations are described in
Kerschbaum et al. (2010).

\begin{figure} 

\begin{minipage}{0.49\textwidth}
\resizebox{\hsize}{!}{\includegraphics{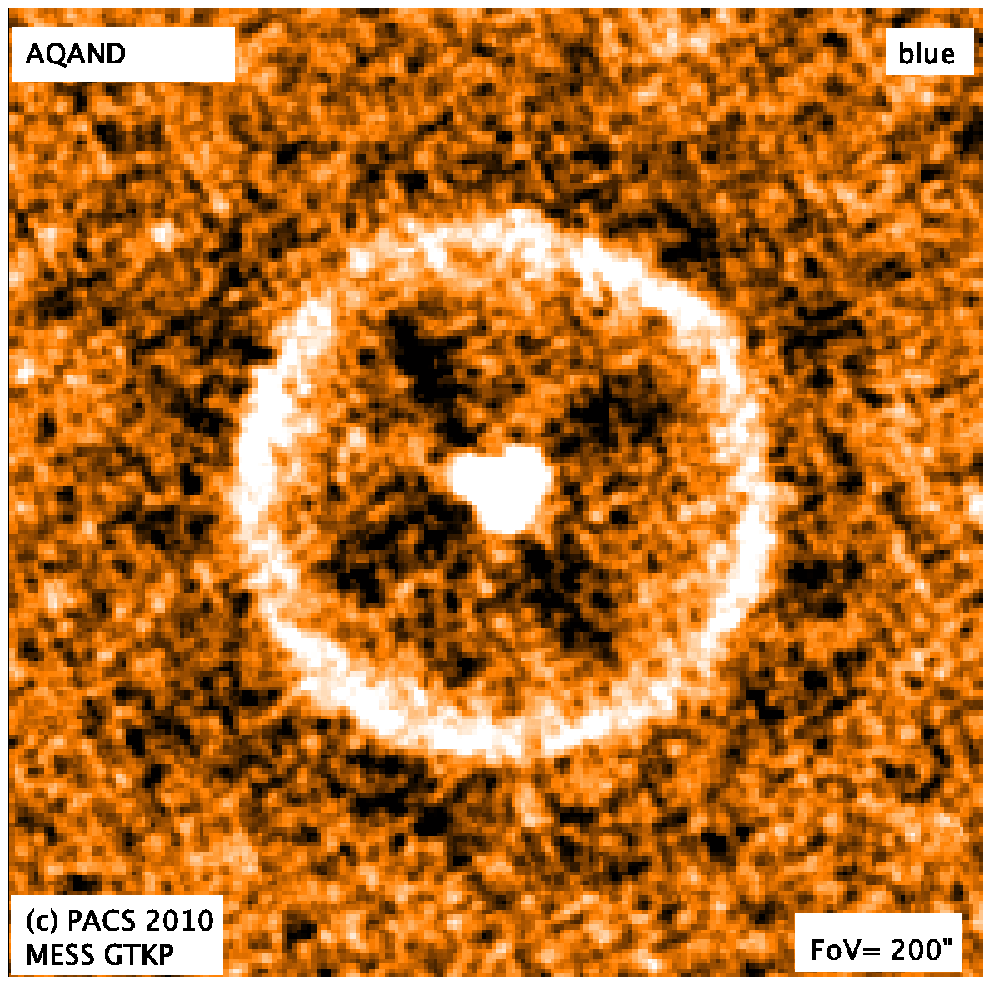}}
\end{minipage}

\begin{minipage}{0.49\textwidth}
\resizebox{\hsize}{!}{\includegraphics{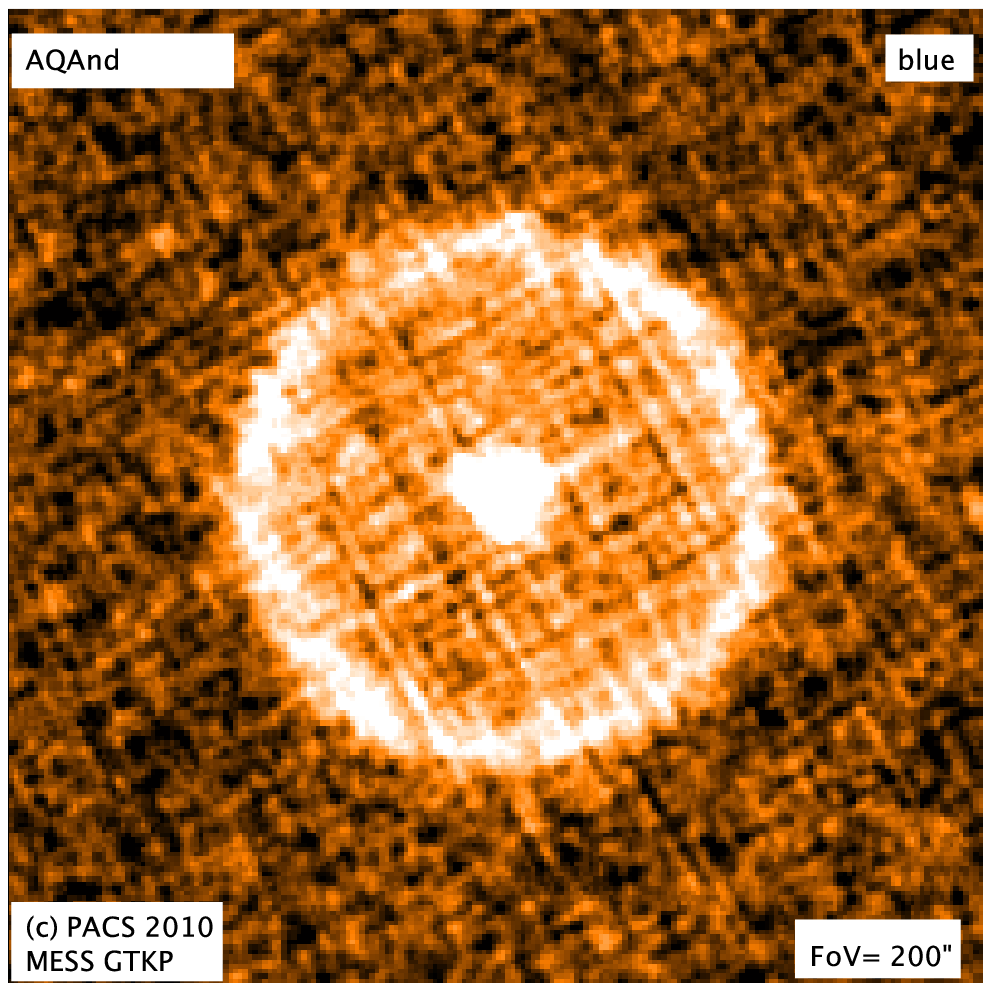}}
\end{minipage}

\caption[]{ 
Top panel:
The final map of AQ And at 70 $\mu$m when the central star and shell are not masked 
during median filtering with a ``high pass filter width'' of 20.
The ``shadowing effect'' is clearly visible.
The central star does not look round but that is due to the secondary mirror support structure which 
introduces a tri-lobe shape of the PSF at a low flux level.
Lower panel:
The map when the central star and shell are masked and a high pass filter width of 35 is used.
} 
\label{Fig-aqand} 
\end{figure}

In earlier versions of HIPE (v1.2) a HPFW as low as 20 was used as default.
The final map (adding scan and cross-scan) with such a low value and no masking is displayed in Fig.~\ref{Fig-aqand}.
The effect of ``shadowing'' that is seen in many level 2 product maps
where there is a bright source in the middle is clearly visible.  
The reason is obvious: without any precaution the median filter will
subtract too much when it passes over locations of significant
emission, in particular a bright central source.  Therefore one sees
dark stripes beside the central source in the two scan
directions, as well as around the detached shell.

\begin{figure} 

\begin{minipage}{0.49\textwidth}
\resizebox{\hsize}{!}{\includegraphics{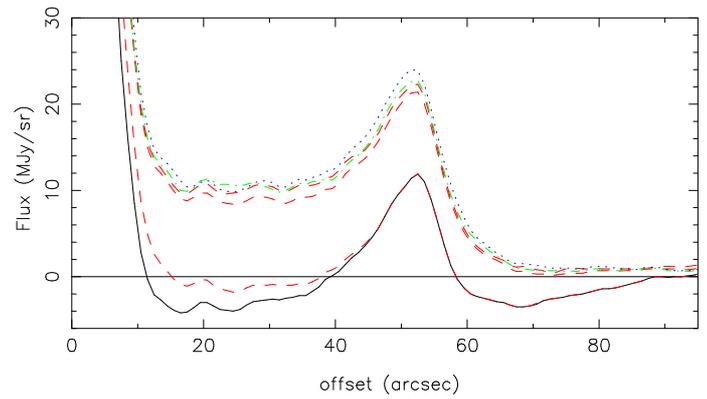}}
\end{minipage}

\caption[]{ 
Azimuthally averaged intensity distribution for AQ And at 70 $\mu$m when
the central star and shell is not masked before the median filtering, 
and using a ``high pass filter width'' (HPFW) of 20 (black solid line),
The red dashed lines indicate the result if a region of 9.8\arcsec\ radius 
around the central star is masked using a  HPFW of 20, 100 and 150 (bottom to top).
The green dashed-dotted line and the blue dotted line show the results if a region 
of radius 60\arcsec, respectively, 80\arcsec\ is masked, with a HPFW of 33, respectively 43.
} 
\label{Fig-1d} 
\end{figure}

The azimuthally averaged 1D-intensity distribution is shown in Fig.~\ref{Fig-1d} 
as the black solid line, illustrating in a different way the regions of negative flux in this map.

The solution to this problem is to mask the central object. 
The task {\tt photReadMaskFromImage} can be used to mask all frames
where the signal exceeds a user given value. In earlier versions of
HIPE however, this threshold was not a free parameter but fixed at a
large value of 0.5 Jy/pixel. Therefore an alternative procedure was developed early on. 
In a first step, the rms, $\sigma$, in the map is determined in a
circle of 30\arcsec\ diameter located $\frac{1}{4}$ of the scan-leg
length to the North of the map centre, 200\arcsec\ in this particular case.
To exclude any possible true sources and undetected glitches in that
area the rms is calculated by first eliminating the lowest and highest
2\% of the data values.

In a second step the contour at a level of $7\sigma$ is determined. 
This is on purpose a large value to ensure a relatively well behaved shape of the contour.
The longest contour near the centre of the map is then approximated by a circle. 
The centre of this circle is taken as the centre of gravity of the contour\footnote{See
e.g. http://local.wasp.uwa.edu.au/$\sim$pbourke/geometry/polyarea/ for an algorithm.}. 
The distances from the centre to all elements of the contour are calculated and 
the radius is taken as the mode (hereafter {\it sourceradius}.).

This radius can then be increased by a user-supplied value. The default 
value is 1.3, which means for most sources that a region around the 
central star down to a level of approximately $3\sigma$ is masked.

In the case of AQ And, the region that is masked has a radius of 9.8\arcsec.  
The effect is shown in Fig.~\ref{Fig-1d} as the red dashed line, for
a HPFW of 20, 100 and 150 (bottom to top). It is obvious that masking
only the central object with a small HPFW does not lead to a reliable
result.  Only for a HPFW of 150 a stable 1D intensity distribution is
reached.  This result is not that surprising as the emission in the
shell has not been masked.  The green dashed-dotted line and the blue
dotted line show the results if a region of radius 60, respectively,
80\arcsec\ is masked. In that case it is not necessary to use a very
large HPFW. In fact, the only concern is that it is large enough 
that the median filter contains unmasked frames\footnote{in particular
  it is calculated from 
  HPFW= max(20, int(1.01$\times$ {\it maskfrac}$\times$ {\it sourceradius}/ scanspeed/ 0.1) +5)), 
  where {\it maskfrac} is a user-supplied number with default value
  1.3. This ensures that even in the worst case the median filter uses
  11 data values to calculate the median. The minimum value of 20 units
  is enforced for cases where the central source is weak and no or no 
  reliable {\it sourceradius} can be determined}.
In the cases shown the HPFW values are 33 (dashed-dotted line) and 43 (dotted line).
The different intensity profiles show that reliable results (i.e. differences in data 
reduction strategies that lead to intensities that are within the photometric 
calibration uncertainly of 10\%) can be achieved if areas in the
map that correspond to significant levels of emission are masked.

In the data reduction scheme employed in the present paper the masked region is where
the signal is above $3\sigma$ in the entire map (using the {\tt photReadMaskFromImage} task), 
and the procedure outlined above to derive a {\it sourceradius}. The
scaling factor of this radius is 1.3 by default to roughly mask a region with a
signal above $3\sigma$ around the central object\footnote{
As clearly demonstrated above, the scaling factor must be larger for
the stars with detached shells or obvious regions of extended emission
in order to compute the appropriate value of the HPFW and ${\it maskfrac}$ factors of
2.0 (U Ant, EP Aqr, o Cet, $\pi$ Gru),
2.5 (TX Psc, X Her),
3.0 ($\theta$ Aps, T Mic, RT Vir),
3.5 ($\alpha$ Ori)
4.0 (R Hya),
5.0 (U Cam, S Cep, RZ Sgr),
6.0 (AQ And, TT Cyg),
8.0 (UX Dra)
were used.  
}.

The {\it sourceradius} is also used in defining the details of the 2nd level deglitching task.
As mentioned above, the {\tt photMMTDeglitching} task incorrectly masks a
significant number of frames due to the brightness and compactness of
the central source. In a first step all frames that are masked by the
photMMTDeglitching task inside a radius 0.5$\times${\it sourceradius} are unmasked.

Then the {\tt IIndLevelDeglitchTask} is used on the map, using  6$\sigma$-clipping over a running box of 41 elements.
The value for the sigma-clipping and the size of the box were
determined by comparing the glitch rate in an off-source area (in fact
the same area as where the rms in the map is calculated) to the
glitch rate in the area around the source for a large number of objects. 
As the radiation events should be distributed randomly the
{\tt photMMTDeglitching} and {\tt IIndLevelDeglitch} tasks should roughly flag the
same percentage of frames.

After the masking of regions of significant emission and the masking
of glitches on source the {\tt highpassFilter} task is run again with the optimised 
value as outlined earlier. 
The frames objects for scan and cross-scan are joined\footnote{using
  the join method on a Frames object: frames.join(frames1)} and the
final map is created using a pixelsize of 1\arcsec\ (blue filter) and 2\arcsec\ (red filter).

\end{appendix}

\end{document}